\documentclass[a4paper,11pt]{article}
\pdfoutput=1 

\usepackage{jheppub} 

\usepackage{graphicx}
\usepackage{amsfonts,amssymb}
\usepackage{color}

\def\Dsl{\hbox{/\kern-.6000em D}} 

\def\dsl{\,\raise.15ex\hbox{/}\mkern-13.5mu D}
\def\bsigma{\mbox{\boldmath $\sigma$}}

\def\bsigma{\mbox{\boldmath $\sigma$}}

\def\ltap{\ \raise.3ex\hbox{$<$\kern-.75em\lower1ex\hbox{$\sim$}}\ }
\def\gtap{\ \raise.3ex\hbox{$>$\kern-.75em\lower1ex\hbox{$\sim$}}\ }
\def\OMIT#1{}

\def\lsim{\mathrel{\raise.3ex\hbox{$<$\kern-.75em\lower1ex\hbox{$\sim$}}}}
\def\gsim{\mathrel{\raise.3ex\hbox{$>$\kern-.75em\lower1ex\hbox{$\sim$}}}}

\def\msb{{\overline{\rm MS}}}

\newcommand{\nn}{\nonumber}

\newcommand{\bmk}{\mathbf k}
\newcommand{\bmp}{\mathbf p}

\newcommand{\bmJ}{\mathbf J}

\newcommand{\bmS}{\mathbf S}

\newcommand{\bmpp}{{\bmp^\prime}}

\newcommand{\ord}{{\cal O}}
\newcommand{\as}{\alpha_S}
\newcommand{\au}{\alpha_U}
\newcommand{\ah}{\alpha_h}

\def\msb{{\overline{\rm MS}}}

\catcode`\@=11
\def\slash{\mathpalette\make@slash}
\def\make@slash#1#2{\setbox\z@\hbox{$#1#2$}%
  \hbox to 0pt{\hss$#1/$\hss\kern-\wd0}\box0}
\catcode`\@=12 

\newcommand{\rot}{\color{red}}
\newcommand{\blau}{\color{blue}}

\definecolor{hellblau}{rgb}{0.5,0.5,1}
\definecolor{newblue}{rgb}{0.2,0.2,0.6}
\definecolor{orange}{rgb}{1,0.5,0}
\definecolor{lila}{rgb}{0.5,0,0.5}
\definecolor{brown}{rgb}{0.6,0.4,0.2}
\newcommand{\orange}{\color{orange}}
\newcommand{\lila}{\color{lila}}
\newcommand{\newblue}{\color{newblue}}
\newcommand{\hellblau}{\color{hellblau}}

\preprint{ \vbox{ 
\hbox{UWThPh-2013-23}
\hbox{DESY 13-168}
}}

\title{The Top-Antitop Threshold at the ILC:\\
NNLL QCD Uncertainties}

\author[a]{Andr\'e~H.~Hoang,} 
\author[b]{Maximilian~Stahlhofen}

\affiliation[a]{University of Vienna, Faculty of Physics, Boltzmanngasse 5, A-1090 Wien, Austria}
\affiliation[b]{DESY Theory Group, Notkestra\ss e 85, D-22607 Hamburg, Germany}

\emailAdd{andre.hoang@univie.ac.at}
\emailAdd{maximilian.stahlhofen@desy.de}

\abstract{
We discuss the top-antitop production cross section near threshold at a future linear collider accounting for the NNLL QCD corrections to the anomalous dimension of the leading S-wave production current computed recently within renormalization-group-improved NRQCD perturbation theory.
We argue that the still unknown soft NNLL mixing corrections are negligible so that the NNLL QCD corrections to the total cross section can be considered complete for practical purposes.
Based on combined variations of the renormalization and matching scales and the overall size of the perturbative corrections we estimate that the NNLL QCD total cross section has a normalization uncertainty of $d \sigma/\sigma=5\%$ at threshold.
We present results for the total cross section and also for the experimentally more relevant case, when moderate cuts are imposed on the reconstructed top and antitop invariant masses. 
}

\begin{document} 
\maketitle
\flushbottom

\section{Introduction}
\label{sectionintro}

The measurement of the top-antitop resonance line-shape is a major goal at a future linear collider (LC)~\cite{Baer:2013cma}.
It will allow for a precise determination of the top quark mass with unambiguous control over the renormalization scheme as well as the width and the couplings of the top quark and therefore provides crucial input for tests of the Standard Model (SM). 
A precise and unambiguous measurement of the top quark mass is also important for vacuum stability studies and the analyses of many new physics models.
In particular the c.m. energy, where the cross section starts to rise, i.e.\ the position of the resonance peak, is very sensitive to the top quark mass. Experimental studies on the scan of the cross section line-shape~\cite{Martinez:2002st,Seidel:2013sqa} including precise knowledge of the $e^+e^-$ luminosity spectrum have shown that a top mass experimental uncertainty well below $100$ MeV is feasible. On the other hand, other parameters like the strong coupling $\alpha_s$, the decay width of the top $\Gamma_t$ or the top Yukawa coupling can be extracted from the normalization and the shape of the cross section in the peak region.
The expected accuracy of the experimental data from a threshold scan requires a very precise theoretical prediction for the cross section in the resonance region. Since nonperturbative effects in top-antitop production are sufficiently suppressed due to the large top quark width~\cite{Fadin:1987wz}, high-order perturbative calculations using nonrelativistic effective field theories (EFT's) based on NRQCD~\cite{Bodwin:1992ye} can in principle yield theoretical uncertainties that can compete with the experimental ones. 
Nevertheless, reaching predictions with small theoretical uncertainties has proven to be a nontrivial task~\cite{Hoang:2000yr}.

The bound-state like dynamics of the top-antitop system close to threshold is governed by (at least) three physical scales: the mass $m$, the 3-momentum $\sim m v$ and the kinetic energy $\sim m v^2$ of the top quarks in the c.m.\ frame. 
They are referred to as the hard, the soft and the ultrasoft scale, respectively.
For the typical velocities in the resonance region ($v\sim 0.1-0.2$) we are therefore confronted with a multi-scale problem
and the top-antitop system becomes sensitive to scales (soft, ultrasoft) that are substantially lower than the top mass.
In NRQCD the strong hierarchy among the physical scales is exploited in order to resum 'Coulomb-singular' terms $\sim (\alpha_s/v)^n$ to all orders in perturbation theory using a Schr\"odinger equation. This resummation is crucial to correctly describe the behavior of the cross section at and near threshold. Higher order corrections (in $\alpha_s$ and $v$) to the leading order (LO) Coulomb resummed result are obtained using nonrelativistic perturbation theory. 
Calculations that account for the resummation of the Coulomb-singular terms in the context of perturbative NRQCD are commonly called ``fixed-order'' computations.

At the next-to-next-to-leading order (NNLO) level such fixed-order calculations were achieved by a number of groups~\cite{Hoang:1998xf,Hoang:1999zc,Melnikov:1998pr,Yakovlev:1998ke,Beneke:1999qg,Nagano:1999nw,Penin:1998mx}.
In Ref.~\cite{Hoang:2000yr} a common effort was made to estimate the uncertainties of the NNLO total threshold cross section. 
The position of the visible top-antitop 1S bound state resonance is very stable when short-distance threshold mass schemes are employed for the top quark in order to avoid an intrinsic renormalon ambiguity of $\ord(\Lambda_{\rm QCD})$ in the perturbative expansion.
This allows for theoretical uncertainties in the top quark mass determinations at the 100 MeV level, which matches well with the expected experimental error.
On the other hand, it was estimated that the normalization of the threshold cross section has uncomfortably large uncertainties at the level of $\pm$20\%, which would make precise measurements of the top width and couplings impossible. To match the expected experimental precision at a future LC~\cite{Martinez:2002st,Seidel:2013sqa} a normalization uncertainty of around 3\% should be achieved in the theoretical predictions.
By now fixed-order partial results have been obtained at N$^3$LO, see e.g.\ Refs.~\cite{Kniehl:2002yv,Beneke:2007gj,Beneke:2008cr,Anzai:2009tm,Smirnov:2009fh} and we refer to Ref.~\cite{Beneke:2008ec} for a preliminary analysis.

Some time ago extensions of the NRQCD formalism have been devised, notably pNRQCD~\cite{pNRQCDfirst,Brambilla:1999xf} and vNRQCD~\cite{Luke:1999kz,Manohar:1999xd,Hoang:2002yy}, which in addition to the Coulomb-singular terms allow the resummation of logarithms of the heavy quark velocity ($\ln v$) related to ratios of the hard, soft and ultrasoft scales. 
Predictions that systematically account for the summation of velocity logarithms are called ``renormalization-group-improved'' (RGI) calculations and may stabilize the normalization of the top threshold cross section.\footnote{RGI and fixed-order approaches yield similar results concerning theoretical uncertainties in the determination of the top quark mass in proper short-distance mass schemes, because these schemes involve infrared subtractions to directly stabilize the 1S ground state top-antitop binding energy.
In the RGI and the fixed-order approaches these subtractions are carried out consistently w.r.t. the treatment of the higher order velocity logarithms and thus lead to comparable stabilizations of the observable ground state peak energy.
}
In this paper we work in the vNRQCD framework, which features a modified renormalization group (RG) with a ``subtraction velocity'' $\nu$ that parametrizes the correlation between the soft renormalization scale, $\mu_S=\nu\, \mu_h$ and the ultrasoft renormalization scale, $\mu_U=\nu^2 \mu_h$ (reflecting the kinematic correlation between the nonrelativistic kinetic energy and 3-momentum), where $\mu_h\sim m$ is the hard scale at which the matching of the EFT to full QCD is performed. In the following we denote the strong coupling $\alpha_s$ at the hard, the soft and the ultrasoft scale as $\ah \equiv \alpha_s(\mu_h)$, $\as \equiv \alpha_s(\mu_S)$ and $\au \equiv \alpha_s(\mu_U)$, respectively, whenever the distinction is necessary.

Upon resummation of the singular terms $\propto (\alpha_s/v)^n$ and $\propto (\alpha_s \ln v)^n$ the RGI R-ratio close to the top-antitop threshold schematically takes the form ($\sigma_{\mu^+\!\mu^-}=4\pi\alpha^2/(3s)$)
\begin{align}
R = \frac{\sigma_{t \bar t}}{\sigma_{\mu^+\!\mu^-}} = v \sum\limits_k \bigg(\!\frac{\alpha_s}{v}\!\bigg)^{\!\!k} \sum\limits_i (\alpha_s\,\ln\, v )^i \times \bigg\{\! 1 \, ({\rm LL});\;\alpha_s, v \, ({\rm NLL});\;\alpha_s^2,\, \alpha_s v,\, v^2\,({\rm NNLL});\ldots \! \bigg\},
\label{Rstruc}
\end{align}
where we adopt the counting $v\sim\alpha_s \ll 1$ and $\alpha_s \ln v \sim 1$ to comply with the Coulombic bound-state-like nature of the resonance~\cite{Beneke:1999qg,Hoang:2000ib,Hoang:2001mm} and indicated the terms belonging to leading-logarithmic (LL) order, next-to-leading logarithmic (NLL) order, etc..

Partial results for the top-antitop total threshold cross section in RGI perturbation theory at NNLL order have already been presented in Refs.~\cite{Hoang:2000ib,Hoang:2001mm,Pineda:2006ri} some time ago.
Their results suffered from the missing NNLL anomalous dimension of the dominant S-wave top pair production current, which at that time was believed to be small. Within this approximation Refs.~\cite{Hoang:2000ib,Hoang:2001mm} estimated 2-3\% theoretical uncertainties for the cross section normalization based on variations of the subtraction velocity parameter $\nu$.
The analysis of Ref.~\cite{Pineda:2006ri} confirmed the findings, but pointed out that the variation associated with the matching scale $\mu_h$ is significantly larger and that an uncertainty of 10\% should be assigned to this partial NNLL order prediction.
Subsequently, a substantial amount of work was invested in computing the missing NNLL corrections to the anomalous dimension of the dominant S-wave production current.
These NNLL corrections consist of two contributions, the non-mixing corrections related to the UV-divergences of three-loop vNRQCD vertex diagrams and the mixing corrections related to the NLL evolution of the potential coefficients entering the NLL anomalous dimension of the current.
The calculation of the complete set of non-mixing contributions\footnote{Expanded in fixed-order the velocity logarithms summed by the NNLL non-mixing corrections appear at N$^3$LO and beyond. At N$^3$LO they arise as single logarithmic terms.} from all three-loop (soft and ultrasoft) vertex diagrams~\cite{Hoang:2003ns} revealed that the non-mixing corrections from ultrasoft gluon diagrams are extremely large, questioning the conclusions drawn in Refs.~\cite{Hoang:2000ib,Hoang:2001mm,Pineda:2006ri}, see Ref.~\cite{Hoang:2003xg}.
Recently the full set of ultrasoft mixing corrections from the subleading RG evolution of the potential coefficients~\cite{Hoang:2006ht,Pineda:2011aw,Hoang:2011gy} has become available~\cite{Hoang:2011gy}.\footnote{Expanded in fixed-order the velocity logarithms summed by the ultrasoft NNLL mixing corrections appear at N$^4$LO and beyond. At N$^4$LO they arise as double logarithmic terms.}
The results show that the ultrasoft mixing and non-mixing NNLL corrections to the evolution of the production current are both anomalously large, but that there are substantial cancellations between them. This appears to render NNLL predictions for the top pair threshold cross section with QCD uncertainties at the level of a few percent feasible~\cite{Hoang:2011it}.

Up to now the full set of all available NNLL QCD corrections have not been applied to the top threshold cross section. It is the aim of this paper to fill this gap and carry out such an updated NNLL analysis with the intention to assess the QCD uncertainties including the effects of combined variations of the soft and ultrasoft renormalization scales $\mu_S$ and $\mu_U$ as well as the matching scale $\mu_h$. 
Currently the only missing contributions for a complete NNLL QCD prediction are the NNLL mixing corrections from soft and potential diagrams contributing to the NLL evolution of the subleading potentials.
We argue that the uncertainty due to these unknown soft mixing corrections is negligible (see also Refs.~\cite{Hoang:2011it,Hoang:2012us}) so that the QCD predictions at the NNLL level can be considered complete from the practical point of view.
We find that accounting for the recently determined ultrasoft NNLL mixing corrections does indeed lead to a substantial stabilization of the results found in Ref.~\cite{Hoang:2003ns}. Moreover, we also observe that the large matching scale dependence reported on in Ref.~\cite{Pineda:2006ri} is reduced considerably. 
Overall we find that the NNLL normalization uncertainty in the top threshold cross section in the RGI approach from QCD effects is $d \sigma/\sigma=5\%$, and that this estimate is fully consistent with the size of the corrections. This closely approaches the theoretical normalization uncertainty of 3\% aimed for in the theoretical predictions.

It has been pointed out in Ref.~\cite{Hoang:2010gu} that the pure NRQCD total cross section contains a sizable unphysical contribution that is related to the leading order effective theory implementation of the top quark width $\Gamma_t$. 
The width effectively shifts the energy into the complex plane, $E \to E + i\Gamma_t$, prior to taking the absorptive part of the forward scattering amplitude. 
This unphysical contribution arises since using the optical theorem within NRQCD strongly overestimates phase space regions of top decay final states from (anti)top quarks with high virtuality. Thus to achieve a realistic theoretical prediction for the top threshold cross section it is necessary to include electroweak effects beyond the complex energy shift related to the top quark width. 
This also includes contributions from irreducible background processes which have the same final state as top pair production~\cite{Hoang:2004tg,Hoang:2006pd,Hoang:2010gu,Beneke:2010mp,Penin:2011gg,Jantzen:2013gpa}.
It was demonstrated in Ref.~\cite{Hoang:2010gu} that the by far largest of these electroweak effects is connected to the implementation of phase space cuts on the top and antitop decay products, which in fact remove the unphysical phase space contributions. 
It was in particular shown that imposing moderate invariant mass cuts on the pure (NR)QCD prediction with the complex energy shift gives a much more realistic description of the total cross section near threshold.

Thus besides studying the total NRQCD cross section based on the optical theorem we will also examine the more realistic case, when loose invariant mass cuts are applied to the decay products of both, the top and the antitop.
Concerning the other electroweak effects we only account for the decay of the top quark at leading order through the complex energy shift.
We refer to Refs.~\cite{Hoang:2004tg,Hoang:2006pd,Hoang:2010gu,Beneke:2010mp,Penin:2011gg,Jantzen:2013gpa} for a more systematic discussion of higher order electroweak corrections. 
It is straightforward to combine them with our results, and they can therefore be studied independently. 
For the  prediction of the inclusive (total) cross section the dominant theoretical error anyway originates from perturbative QCD effects.
Hence our treatment of electroweak effects is sufficient for the purpose of this work.
Overall we find that the behavior of the QCD effects and in particular the QCD uncertainties of the NNLL cross section are not affected by phase space cuts.
We finally note that we consider the 1S mass~\cite{Hoang:1999zc} defined as half the perturbative (QCD) contribution to the $^3S_1$ ground state of would-be toponium as the short-distance mass in our analysis. For alternative threshold schemes suggested in the literature we refer to Refs.~\cite{Beneke:1998rk,Hoang:1999zc,Pineda:2001zq}.

The content of this work is as follows:
In Sec.~\ref{sectioncrosssection} we briefly review the vNRQCD calculation of the RGI total $t \bar t$ threshold cross section and summarize the results for the different contributions to NNLL order. We also argue that the present status of the calculation allows for a very good approximation of the full NNLL cross section and that the error related to the missing NNLL terms is negligible compared to the perturbative uncertainty of the complete NNLL result.
In Sec.~\ref{sectionanalysis} we perform combined variations of the renormalization and matching scales in the NNLL vNRQCD expression for the total inclusive cross section and compare the resulting uncertainties to previous analyses.
Section~\ref{sectioncuts} addresses the effect of invariant mass cuts for the decay products of the top and antitop quark and gives our final estimate for the overall perturbative uncertainty of the physical cross section at NNLL.
We conclude in Sec.~\ref{sectionconclusion}.

\section{The NNLL Total Cross Section}
\label{sectioncrosssection}

In the following we briefly review the theory setup for the vNRQCD prediction of the top-antitop threshold cross section at NNLL order concentrating mainly on the new NNLL mixing corrections to the anomalous dimension of the leading order S-wave current~\cite{Hoang:2011gy,Hoang:2011it,Hoang:2012us}. For details on the other theoretical input we refer to Refs.~\cite{Hoang:2000ib,Hoang:2001mm,Hoang:2002yy,Hoang:2003ns}.

We consider the production of the top-antitop pair in $e^+ e^-$ collisions mediated by a virtual photon or Z boson with the c.m. energy $\sqrt{s}$.
The R-Ratio for the total cross section therefore has vector and axial-vector contributions:
\begin{align}
R^{\gamma,Z}(s) = \sigma_{\rm tot}(e^+ e^- \to \gamma^*,Z^* \to t\,{\bar t}\,)/\sigma_{\mu^+\!\mu^-}= F^v(s)\,R^v(s) +  F^a(s) R^a(s) \,. \label{Rtot}
\end{align}
Employing the optical theorem the $R^{v,a}$ can be related to current-current correlators, 
\begin{align}
 R^v(s) &= \frac{4 \pi }{s}\,\mbox{Im}\,\left[-i\int d^4x\: e^{i\sqrt{s}\,t }
  \left\langle\,0\,\left|\, T\, j^v_{\mu}(x) \,
  {j^v}^{\mu} (0)\, \right|\,0\,\right\rangle\,\right] \,, \nn\\
 R^a(s) &= \frac{4 \pi }{s}\,\mbox{Im}\,\left[-i\int d^4x\: e^{i\sqrt{s}\,t }
  \left\langle\,0\,\left|\, T\, j^a_{\mu}(x) \,
  {j^a}^{\mu} (0)\, \right|\,0\,\right\rangle\,\right] \,.
\label{Rratios}
\end{align}
The respective prefactors $F^v(s)$, $F^a(s)$ account for the (tree-level) $\gamma$ and Z exchange and are given e.g. in Ref.~\cite{Hoang:2001mm}. The Standard Model (SM) currents $j^v_{\mu}$ and $j^a_{\mu}$ produce the heavy quark pair in a vector and an axial-vector state, respectively. In the effective theory these currents are replaced by their nonrelativistic counterparts through an operator product expansion and we find to NNLO in the $v$ counting
\begin{align}
R^v(s) &= \frac{4\pi}{s}\,
 \mbox{Im}\Big[\,
 c_1^2(\nu,h)\,{\cal A}_1(v,m,\nu,h) + 
 2\,c_1(\nu,h)\,c_2(\nu,h)\,{\cal A}_2(v,m,\nu,h) \,\Big] \,, 
\nn\\
 R^a(s) &=  \frac{4\pi}{s}\,
 \mbox{Im}\Big[\,c_3^2(\nu,h)\,{\cal A}_3(v,m,\nu,h)\,\Big]\,.
\label{effRratios}
\end{align}
For later reference we not only indicate the renormalization scale dependence of the various $c_i$ and ${\cal A}_i$ terms in Eq.~\eqref{effRratios} through the RG parameter $\nu$ but also their dependence (explicit or implicit) on the matching scale through the matching parameter $h\equiv \mu_h/m$. Both $\nu$ and $h$ are dimensionless. Throughout this paper we are employing the $\msb$ scheme for renormalization (and matching). In the remainder of this section we discuss the individual terms in Eq.~\eqref{effRratios} in some detail.

The effective current correlators
\begin{align}
{\cal A}_i(v, m, \nu,h) &= i\,
 \sum\limits_{\bmp,\bmpp}
 \int\! d^4x\: e^{i (\sqrt{s} - 2 m) t}\:
 \Big\langle\,0\,\Big|\, T\, \bmJ_{i,\bmp}(x) \, \bmJ^\dagger_{i,\bmpp}(0)
 \Big|\,0\,\Big\rangle \,, 
\label{A1}
\end{align}
have a well-defined scaling in the nonrelativistic velocity $v$.
In Eq.~\eqref{effRratios} for instance ${\cal A}_2$ and ${\cal A}_3$ are $v^2$-suppressed compared to the LO correlator ${\cal A}_1$.
The correlator ${\cal A}_1$ is determined from the S-wave zero-distance Green function of the Schr\"odinger equation for the top-antitop system. Following Refs.~\cite{Hoang:2000ib,Hoang:2001mm} we use a semi-analytic approach to calculate ${\cal A}_1$, where all effects from the Coulomb potential (including its corrections up to NNLL order, i.e. $\ord(\as^3)$), are accounted for exactly through a numerical solution for the Green function using the \mbox{TOPPIK} program~\cite{Hoang:1999zc} and all $\ord(v^2)$ corrections are determined analytically using the results from Ref.~\cite{Hoang:2001mm} with the updates concerning the convention for the $1/(m\bmk)$ potential given in Refs.~\cite{Hoang:2002yy,Hoang:2003ns}. For the $v^2$-suppressed correlators ${\cal A}_{2,3}$ we employ the analytic results given in Ref.~\cite{Hoang:2001mm}.

The $c_i$ in Eq.~\eqref{effRratios} are the Wilson coefficients in the nonrelativistic expansion of the vector and axial-vector SM currents in Eq.~\eqref{Rratios}. The results for $c_2$ and $c_3$ have been given analytically in Ref.~\cite{Hoang:2001mm}. Since they multiply the $v^2$-suppressed correlators ${\cal A}_{2,3}$, the numerical impact of their anomalous dimensions is quite small. 
Their contribution is not significant for the perturbative uncertainties, but included of course for completeness.
The term $c_1$ is the Wilson coefficient of the leading $^3S_1$ effective current
\begin{align} 
  {\bmJ}_{1,\bmp} = 
    \psi_{\bmp}^\dagger\, \bsigma (i\sigma_2) \chi_{-\bmp}^*
\,,
\label{J1J0}
\end{align}
where the field operators $\psi_{\bmp}$ and $\chi_{\bmp}$ (in the vNRQCD label notation, with suppressed color indices) annihilate top and antitop quarks with soft three-momentum $\bmp$, respectively. At the LL order $c_1$ is renormalization scale invariant. The NNLL order anomalous dimension contains the previously mentioned mixing and non-mixing corrections according to Refs.~\cite{Hoang:2011gy,Hoang:2011it} and Ref.~\cite{Hoang:2003ns}, respectively.

When electroweak effects and the top quark decay are accounted for beyond the leading order level, the top pair threshold cross section has to be defined through the top decay final state and the form of the Eqs.~\eqref{Rratios} and~\eqref{effRratios} needs to be extended due to interference contributions involving non-$t\bar t$ amplitudes and irreducible background. 
Moreover, the cross section becomes intrinsically dependent on cuts applied on the $t \bar t$ final state. For details we refer to Refs.~\cite{Hoang:2004tg,Hoang:2006pd,Hoang:2010gu}. 
As mentioned above, the aim of this work is to study the perturbative uncertainties from QCD effects at the NNLL order level. Concerning the electroweak effects we account (i) for the top quark on-shell width and (ii) for cuts on the reconstructed top and antitop invariant masses. The top quark width is implemented in the optical theorem relation Eq.~\eqref{effRratios} by using the common complex definition
\begin{align}
v = \sqrt{\frac{\sqrt{s}-2m + i\Gamma_t}{m}}
\label{veff}
\end{align}
for the effective velocity. This approach to implement the top quark width is known to fully account for electroweak effects at the leading order level~\cite{Fadin:1987wz} (using the counting $v^2\sim \alpha_s^2\sim \alpha_{\rm ew}$) but also entails that the QCD cross section based on Eq.~\eqref{effRratios} receives the sizable contributions from unphysical phase space regions associated with arbitrary large top and antitop invariant masses. The problem is related to the nonrelativistic expansion in the large top quark mass and leads to an unphysical enhancement of the total cross section in Eq.~\eqref{effRratios} that, formally, acts like a background contribution. In a full treatment of electroweak effects, the form of Eq.~\eqref{effRratios} has to be extended by additive contributions from local $(e^+e^-)(e^+e^-)$ forward scattering operators, and all Wilson coefficients in general receive complex contributions through electroweak matching conditions~\cite{Hoang:2004tg,Hoang:2006pd,Hoang:2010gu}, see in particular Eq.~(26) in Ref.~\cite{Hoang:2010gu}. 

It has been demonstrated in Ref.~\cite{Hoang:2010gu} that the unphysical phase space contributions just mentioned (and the corrections related to removing them) contribute at the next-to-leading order level and constitute the by far largest numerical electroweak matching effects. 
It has also been shown how these unphysical phase space contributions arise within the NRQCD effective theory diagrams and, that they can be computed by imposing a cut $\Delta M$ on the top and antitop invariant masses in NRQCD phase space integrations:
\begin{align}
(m-\Delta M) \le M_{t,\bar t} \le (m+\Delta M)\,,
\label{pscut}
\end{align}
where $M_{t,\bar t}$ is the (anti)top reconstructed invariant mass. 
In fact, genuine electroweak effects related to background contributions are so much smaller that the NRQCD cross section based on Eq.~\eqref{effRratios} with NRQCD cuts on the top and antitop invariant masses represents a much better approximation to the physical total cross section than Eq.~\eqref{effRratios} when only the width effect (i) is accounted for~\cite{Hoang:2010gu}.  
We note that instead of employing the full set of analytic phase space matching conditions for the $(e^+e^-)(e^+e^-)$ forward scattering operators discussed in detail in Ref.~\cite{Hoang:2010gu}, we use these analytic results in our analysis only for ${\cal A}_2$ and ${\cal A}_3$ as well as for the $\ord(v^2)$ suppressed contributions in ${\cal A}_1$.
The corresponding terms are given in Eqs.~(54),~(55),~(71) in Ref.~\cite{Hoang:2010gu}, where we do not account for the interference contributions and time dilatation corrections related to the top width.
For the Coulomb interaction contributions in ${\cal A}_1^c$ we implement the invariant mass cuts numerically using an explicit phase space integration for the $t \bar t$ final state and the exact numerical result for the square of the top quark 3-momentum distribution, see Eqs. (79) and (44), (46) in Ref.~\cite{Hoang:2010gu}. 
The 3-momentum distribution is also computed by the \mbox{TOPPIK} routine and we refer to Ref.~\cite{Hoang:1999zc} for details. Since we ignore all other electroweak and top width related corrections and in particular also the anomalous dimensions of the $(e^+e^-)(e^+e^-)$ forward scattering operators our results do not account for a systematic summation of logarithms (of $v$) multiplied by the electroweak couplings. 
Note that we calculate the effects from the invariant mass cuts in the ${\cal A}_i$ correlators with all couplings and potential coefficients at the low energy scales of the correlators. So for the purpose of our analysis we do not treat the contributions from the invariant mass cuts as hard corrections.

For the implementation of the 1S top quark mass scheme its relation to the pole mass scheme is required. At NNLL order it reads
\begin{align}
M^{\rm pole} &= M^{1S} \{1+\Delta^{\rm LL} + \Delta^{\rm NLL} + [(\Delta^{\rm LL})^2 + \Delta_c^{\rm NNLL} + \Delta_m^{\rm NNLL}]\}\,,
\label{mpoletom1s}
\end{align}
where the various $\Delta$ terms are given in Ref.~\cite{Hoang:2001mm} and the additional modifications due to keeping $h \neq 1$ as a free parameter are described in Ref.~\cite{Hoang:2012us}.
The terms $\Delta^{\rm LL,NLL}$ and $\Delta_c^{\rm NNLL}$ arise from the Coulomb potential (including its subleading corrections) and $\Delta_m^{\rm NNLL}$ refers to the relativistic $\ord(v^2)$-term from all other non-Coulomb interactions and kinematic corrections.
As already explained in Ref.~\cite{Hoang:2001mm} 
$\Delta^{\rm LL,NLL}$ and $\Delta_c^{\rm NNLL}$ are implemented exactly within the numerical solution for the contribution to ${\cal A}_1$ arising from the Coulomb interactions.
The corrections due to $\Delta_m^{\rm NNLL}$ are consistently treated perturbatively and give an analytic $\ord(v^2)$ contribution to ${\cal A}_1$. 
In the $v^2$-suppressed correlators ${\cal A}_{2,3}$ only the LL term $\Delta^{\rm LL}$ is required.
In the following we use the notation $m \equiv M^{1S}$.

Since we want to study also variations of the matching scale we retain the explicit $h$ dependence of the terms in Eq.~\eqref{effRratios}. While in the earlier literature often only expressions for $h=1$ can be found, the nontrivial $h$-dependence in the current coefficient $c_1(\nu,h)$ has been worked out in Ref.~\cite{Hoang:2012us}. 
To NNLL the current correlators ${\cal A}_i$ correspond to vNRQCD matrix elements with only soft and potential interactions. Therefore their only explicit dependence on renormalization/matching scales is through $\mu_S$.
Thus the explicit $h$-dependence of the ${\cal A}_i$'s is trivially obtained by replacing $\nu\to h \nu$ in the results for $h=1$ as given in Ref.~\cite{Hoang:2001mm}. 
The coefficients $c_2$ and $c_3$ are only needed at LL. At this order they do not explicitly depend on the matching scale. Implicit $h$-dependence of all contributions to the cross section through the strong coupling constant $\alpha_s$ evaluated at the different renormalization/matching scales ($\ah \equiv \alpha_s(\mu_h)$, $\as \equiv \alpha_s(\mu_S)$ and/or $\au \equiv \alpha_s(\mu_U)$) is understood. This includes in particular the implicit $h$-dependence through the Wilson coefficients ${\cal V}_j(\as,\au)$ of the potentials in the ${\cal A}_i$ correlators.

As outlined above, all terms in the NNLL cross section, Eq.~\eqref{effRratios}, except for $c_1(\nu,h)$, the coefficient of the leading top pair production current, are known to the required precision.
This is also true for the matching condition $c_1(1,h)$, which has been calculated up to two loops~\cite{Czarnecki:1997vz,Beneke:1997jm}.
The most recent progress towards the complete RGI cross section, Eq.~\eqref{Rtot}, at NNLL order has been made in the computation of the RG running of $c_1(\nu,h)$.
To summarize the different contributions to the RG evolution of $c_1$ we parametrize it as
\begin{align}
 \ln\Big[ \frac{c_1(\nu,h)}{c_1(1,h)} \Big] & =
\xi^{\rm NLL}(\nu,h) + 
\Big(\,
\xi^{\rm NNLL}_{\rm m}(\nu,h) + \xi^{\rm NNLL}_{\rm nm}(\nu,h)
\,\Big) + \ldots
\,,
\label{c1solution}
\end{align}
where $\xi^{\rm NLL}$ refers to the NLL order contribution and the $\xi^{\rm
  NNLL}_{\rm m}$ and $\xi^{\rm NNLL}_{\rm nm}$ to the NNLL order mixing and
non-mixing corrections, respectively. The matching condition $c_1(h,1)$ and the
expressions for the $\xi$'s can be found in App.~B of Ref.~\cite{Hoang:2012us} and the references cited there.

Based on the recently completed calculation of the ultrasoft NLL running of the
Wilson coefficients associated to the $\ord(v^2)$- and $\ord(\alpha_s v)$-suppressed 
potentials~\cite{Pineda:2011aw,Hoang:2011gy,Hoang:2006ht}, 
the ultrasoft mixing contributions to $\xi^{\rm NNLL}_{\rm m}$, referred to as
$\xi^{\rm NNLL}_{\rm m, usoft}$, have been determined in Ref.~\cite{Hoang:2011gy}.
Concerning the corresponding soft mixing contributions currently only those coming from
the NLL order anomalous dimension of the spin-dependent $\ord(v^2)$-suppressed
potential are fully known and found to be tiny~\cite{Penin:2004ay}.
On the other hand, because the NLL matching conditions of all
the suppressed potentials are known, it is possible to compute the fixed-order term 
$\propto\alpha_s^3\ln\nu$ of the soft mixing contributions, which we call 
$\xi^{\rm  NNLL}_{\rm m, soft 1}$:
\begin{align}
\xi^{\rm NNLL}_{\rm m, soft1}  =& \frac{\ah^3}{48 \pi}\, C_F^2\, 
\bigg[ C_A \,\Big(16\,\bmS^2-3\Big) 
     + 4\, C_F\,\Big(5 - 2\,{\bf S}^2\Big) - \frac{16}{5}\, T_F \bigg] \ln \nu \, \nn\\
&+ \frac{\ah^3}{12 \pi} C_F^2\,\bigg[ C_A \,\Big(7 \,\bmS^2-17\Big) - 2\, C_F \bigg]\ln h\, \ln \nu\ .
\label{softmixlog}
\end{align}
The result in Eq.~\eqref{softmixlog} has already been given in Ref.~\cite{Hoang:2003ns} for the special case $h=1$. We include the $\ln h$ term here for completeness, but its numerical effect is irrelevant. 
For $\nu \sim v$ the soft mixing logarithm $\xi^{\rm NNLL}_{\rm m, soft1}$ is part of the N$^3$LO result in the fixed-order expansion. By including it in our analysis we make sure that we correctly incorporate all logarithmic terms through N$^3$LO: $\xi^{\rm NNLL}_{\rm m, soft} =\xi^{\rm NNLL}_{\rm m, soft1} + \ord(\as^4 \ln^2 \nu)$.

The NNLL order non-mixing contributions in $\xi^{\rm NNLL}_{\rm nm}$, both, the ultrasoft as well as the soft contributions, 
are completely known already from Ref.~\cite{Hoang:2003ns}.\footnote{The non-mixing term $\xi^{\rm NNLL}_{\rm nm}$ is not yet available in the pNRQCD formalism.}
In the same publication it was observed that the
ultrasoft non-mixing contributions are more than an order of magnitude larger
than the soft ones, and that the smallness of the latter was not arising from any
accidental cancellation between different color factors but was a genuine
property of all soft non-mixing contributions.
On the other hand, the large size of the ultrasoft contributions was not only due to the larger ultrasoft coupling $\au>\as$, but also due to a rather large overall coefficient in the ultrasoft NNLL anomalous dimension.
As was shown in Refs.~\cite{Hoang:2011gy,Hoang:2011it,Hoang:2012us} also the ultrasoft mixing corrections $\xi^{\rm NNLL}_{\rm m,usoft}$ are anomalously large. They have the opposite sign w.r.t. $\xi^{\rm NNLL}_{\rm nm,usoft}$ and lead to a significant cancellation. Still the sum $\xi^{\rm NNLL}_{\rm m,usoft}+\xi^{\rm NNLL}_{\rm nm,usoft}$ is about 20 times larger than $\xi^{\rm NNLL}_{\rm nm,soft}$.

Although the complete soft NNLL mixing corrections $\xi^{\rm NNLL}_{\rm m,soft}$ are still unknown, we argue in the following that their contribution is small and in fact negligible in view of the remaining perturbative uncertainty due to the NNLL ultrasoft corrections, $\xi^{\rm NNLL}_{\rm m,usoft}+\xi^{\rm NNLL}_{\rm nm,usoft}$.
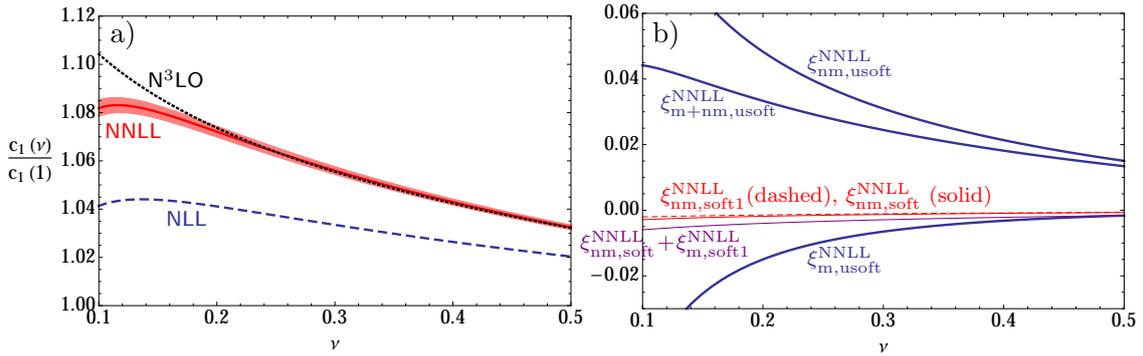
\begin{figure}[ht]
\includegraphics[width=\textwidth]{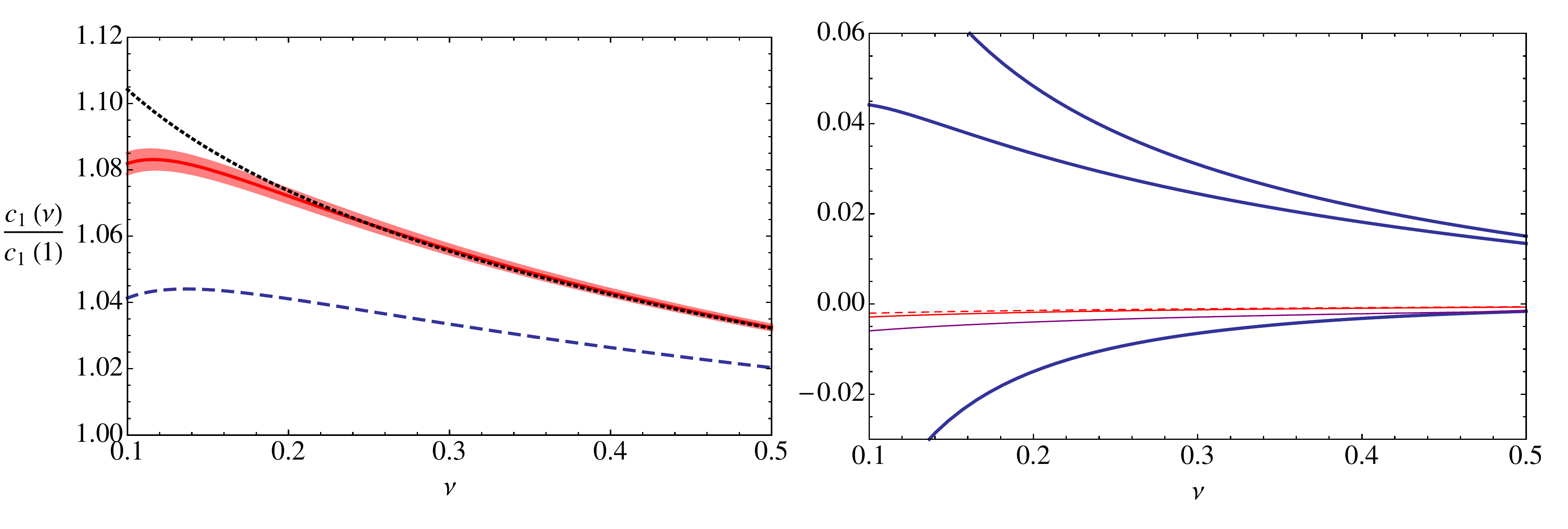}
\put(-377,108){\sf \scriptsize N$^3$LO}
\put(-370,56){\newblue \sf \scriptsize NLL}
\put(-393,89){\rot \sf \scriptsize NNLL}
\put(-391,125){a)}
\put(-188,125){b)}
\put(-130,115){\scriptsize \newblue $\xi_{\rm nm,usoft}^{\rm NNLL}$}
\put(-130,41){\scriptsize \newblue $\xi_{\rm m,usoft}^{\rm NNLL}$}
\put(-185,100){\scriptsize \newblue $\xi_{\rm m+nm,usoft}^{\rm NNLL}$}
\put(-185,65){\scriptsize \rot $\xi_{\rm nm,soft1}^{\rm NNLL}$(dashed), $\xi_{\rm nm,soft}^{\rm NNLL}$ (solid)}
\put(-215,47){\scriptsize \lila $\xi_{\rm nm,soft}^{\rm NNLL}\!+\!\xi_{\rm m,soft1}^{\rm NNLL}$}
\caption{Panel a): 
RG evolution of the  ${}^3S_1$ current coefficient
$c_1(\nu)\equiv c_1(\nu,1)$ normalized to $c_1(1)\equiv c_1(1,1)$ for $m=172$~GeV and $h=1$. 
The blue dashed line represents the full NLL result $\exp(\xi^{\rm NLL})$.
The red solid curve labeled "NNLL" includes in addition all known NNLL corrections: $\exp(\xi^{\rm NLL} + \xi_{\rm nm}^{\rm NNLL}+\xi^{\rm NNLL}_{\rm m, usoft}+\xi^{\rm NNLL}_{\rm m, soft1})$.  
The light red band around the NNLL line is generated by varying the NNLL soft non-mixing
contribution to that curve by a factor between 0 and 2. 
The black dotted line displays the complete N$^3$LO fixed-order expression for $c_1(\nu)/c_1(1)$ given in Eq.~\eqref{c1N3LO}.
Panel b):
Separate curves for the different (soft/ultrasoft, mixing/non-mixing) NNLL
corrections ($\xi^{\rm NNLL}$) to the running of $c_1$ as indicated in the
plot.  
For both plots we have used $\alpha_s^{(n_f=5)}(172~\mbox{GeV})=0.108$.
\label{c1XiPlot}} 
\end{figure}
To estimate the theory error of the NNLL prediction due to the missing soft mixing logarithms in $c_1$ we have plotted the RG running of $c_1$ including all known corrections to NNLL ($h=1$, red solid line) in Fig.~\ref{c1XiPlot}a enclosed in the red band generated by varying the soft non-mixing contributions by a factor between zero and two.
For comparison we also show the NLL RG evolution of $c_1$ (blue dashed line) in the same panel. 
The small numerical size of the soft NNLL non-mixing corrections is evident. In fact, their effect is smaller than the perturbative uncertainty that arises from scale variations of the ultrasoft NNLL corrections in the $\nu$-range between 0.1 and 0.2. We now argue that the complete soft mixing corrections are very likely of similar size as the small NNLL soft non-mixing corrections.

In Fig.~\ref{c1XiPlot}b we have plotted all known soft and ultrasoft mixing as well as non-mixing corrections of NNLL order for $h=1$.
The curve for $\xi^{\rm NNLL}_{\rm nm, soft1}$ denotes the linear logarithmic terms $\alpha_s^3 \ln \nu$ contained in the NNLL soft non-mixing corrections.
We see that $\xi^{\rm NNLL}_{\rm nm, soft1}$ and $\xi^{\rm NNLL}_{\rm m, soft1}$ have very similar size, in fact we have
$\xi^{\rm NNLL}_{\rm m, soft1}/\xi^{\rm NNLL}_{\rm nm, soft1} \approx 1.5$ for $h=1$. 
On the other hand the first NNLL soft non-mixing logarithm already represents the bulk of the resummed result: $0.7<\xi^{\rm NNLL}_{\rm nm, soft1}/\xi^{\rm NNLL}_{\rm nm, soft} <0.9$ for $0.1<\nu<0.5$, as is clearly visible in Fig.~\ref{c1XiPlot}b. Assuming a similar behavior for the NNLL soft mixing logarithms we believe it is safe to argue, that the full NNLL result for $c_1$ should lie well within the error band in Fig.~\ref{c1XiPlot}a. 
The gap between the NLL and the NNLL curve in Fig.~\ref{c1XiPlot}a as well as the residual $\nu$ dependence of the NNLL result is much larger than the width of this band. 
We therefore conclude, as already indicated, that we can safely neglect the uncertainty due to the unknown NNLL soft mixing logarithms in the running of $c_1$ as compared to the residual scale uncertainties that the cross section exhibits due to the NNLL ultrasoft corrections.
The outcome of this consideration remains unchanged for different values of $h$ between one half and two.
In the following we will therefore regard the result for the total top-antitop production cross section in Eqs.~\eqref{Rtot},~\eqref{effRratios} as complete through NNLL.

In Fig.~\ref{c1XiPlot}a we have also added a curve for the fixed-order expansion of $c_1(\nu,1)/c_1(1,1)$ in $\ah$ to N$^3$LO (black dotted line). The corresponding analytic expression is given explicitly in Eq.~\eqref{c1N3LO}. Note that this expression is the complete N$^3$LO fixed-order result for $c_1(\nu,h)/c_1(h)$. Hence the difference of the black dotted line to the solid red one shows the effects of the resummed NNLL logarithms in the current coefficient from beyond N$^3$LO. We see that these higher order contributions are essential for reaching stability in the region $\nu<0.2$, which is crucial for predictions of the top-antitop cross section in the peak region.

\section{Analysis of QCD Uncertainties}
\label{sectionanalysis}

\begin{figure}[t]
\begin{center}
\includegraphics[width=0.35\textwidth]{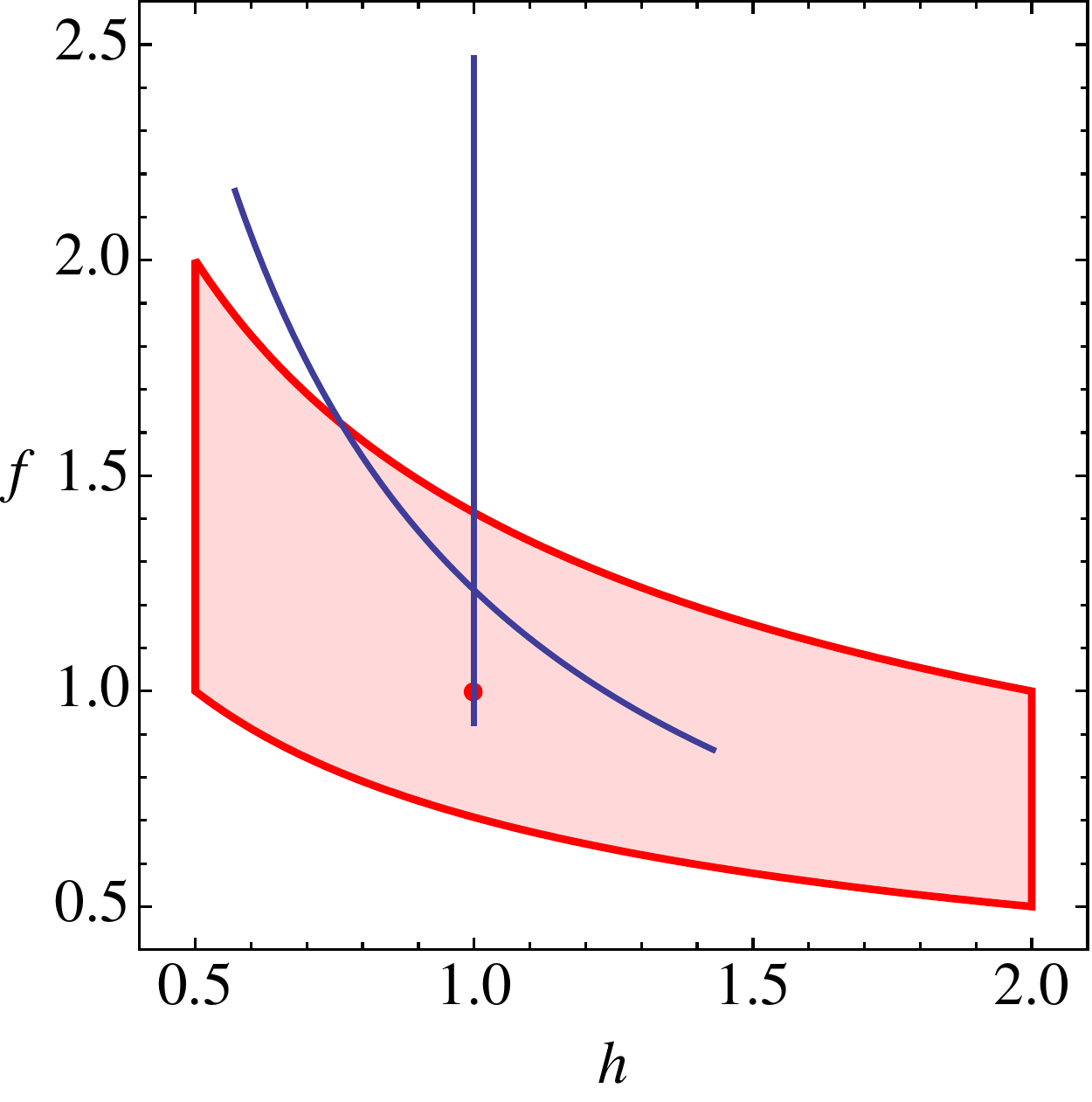}
\end{center}
\caption{Scale variations parametrized by $h=\mu_h/m$ and $f=\nu/\nu_*$. The region bounded by the red line is defined by $1/2\le h f^2 \le2$ and $1/2<h<2$ and represents the parameter space we scan to determine the uncertainty of the cross section due to scale variations. The red dot indicates our default choice for the scale parameters $h$ and $f$. The blue lines correspond to the scale variations studied in Ref.~\cite{Pineda:2006ri} defined w.r.t.\ a different $\nu_*$ definition ($\nu_*=0.185$ compared to our $\nu_*=0.143$ for $\sqrt{s}=2m$). The matching scale variation in Ref.~\cite{Pineda:2006ri} corresponds to a correlated $h$-$f$ variation in our parametrization.
\label{hfregion}}
\end{figure}

In our numerical analysis we allow for variations of the hard matching scale as well as the soft and ultrasoft renormalization scales subject to the following physically motivated constraints:

\begin{itemize}
\item[1.] At all times we maintain the correlation $\mu_U \propto \mu_S^2/m$ and we impose the constraint that at the matching point we have $\mu_S=\mu_U=\mu_h$ such that
the soft and the ultrasoft renormalization scale can never exceed the matching scale $\mu_h$.
\item[2.] We consider variations of the matching scale in the canonical range $m/2 \le \mu_h \le 2 m$.
\item[3.] We consider variations of the ultrasoft scale in the range $\mu_U^*/2 \le \mu_U \le 2 \mu_U^*$, where $\mu_U^*\equiv m \nu_*^2$ is an energy-dependent default choice for the ultrasoft scale with the default subtraction velocity $\nu_*$ being related to the typical velocity.
\end{itemize}
Constraint 1 implies a strict correlation which allows only for a twofold variation.
We therefore parametrize
\begin{align}
\mu_h=h \,m\,,\quad \mu_S=\nu\, h\, m \,,\quad \mu_U=\nu^2 \,h\, m \,,
\label{parametrization}
\end{align}
where $\nu=1$ corresponds to the matching point. For the energy-dependent default subtraction velocity $\nu_*$ we adopt the expression
\begin{align}
\nu_*= 0.05 + \bigg| \sqrt{\frac{\sqrt{s}-2m + i\Gamma_t}{m}}  \bigg|\,.
\label{nudef}
\end{align}
The small constant offset is motivated by the observation from Ref.~\cite{Hoang:2012us} that the typical subtraction velocity scale in the $n$-th moment of the
heavy quark pair threshold cross section can be parametrized very well by the form ${\rm const.} + 1/\sqrt{n}$, where $1/\sqrt{n}$ is of order of the typical velocity. 
We have chosen 0.05 for the offset accounting for the fact that the velocities in the top quark case are substantially smaller than for the bottom quark case that was considered
in Ref.~\cite{Hoang:2012us}.
Together with Eq.~\eqref{nudef} constraint 3 ensures that the ultrasoft scale always remains in the perturbative regime $\mu_U^*/2 > 0.01\,m$. 
In Ref.~\cite{Pineda:2006ri} the choice $\nu_* = 2\, v $ was employed, which leads to similar results in the near-threshold region we consider in this work, but leads to an imbalance of the
scale hierarchies in the intermediate region where the NRQCD cross section might be merged with the full QCD cross section. 
We also define the energy-independent scaling parameter $f=\nu/\nu_*$. 
All scale variations consistent with the constraints 1-3 can then be conveniently translated into variations 
of the variables $f$ and $h$ with the restrictions
\begin{align}
1/2\le h f^2 \le 2\,, \quad 1/2<h<2\,.
\end{align}
The corresponding region in the two-dimensional $h$-$f$ plane is illustrated by the red area in Fig.~\ref{hfregion}.  

\begin{figure}[ht]
\includegraphics[width=0.49\textwidth]{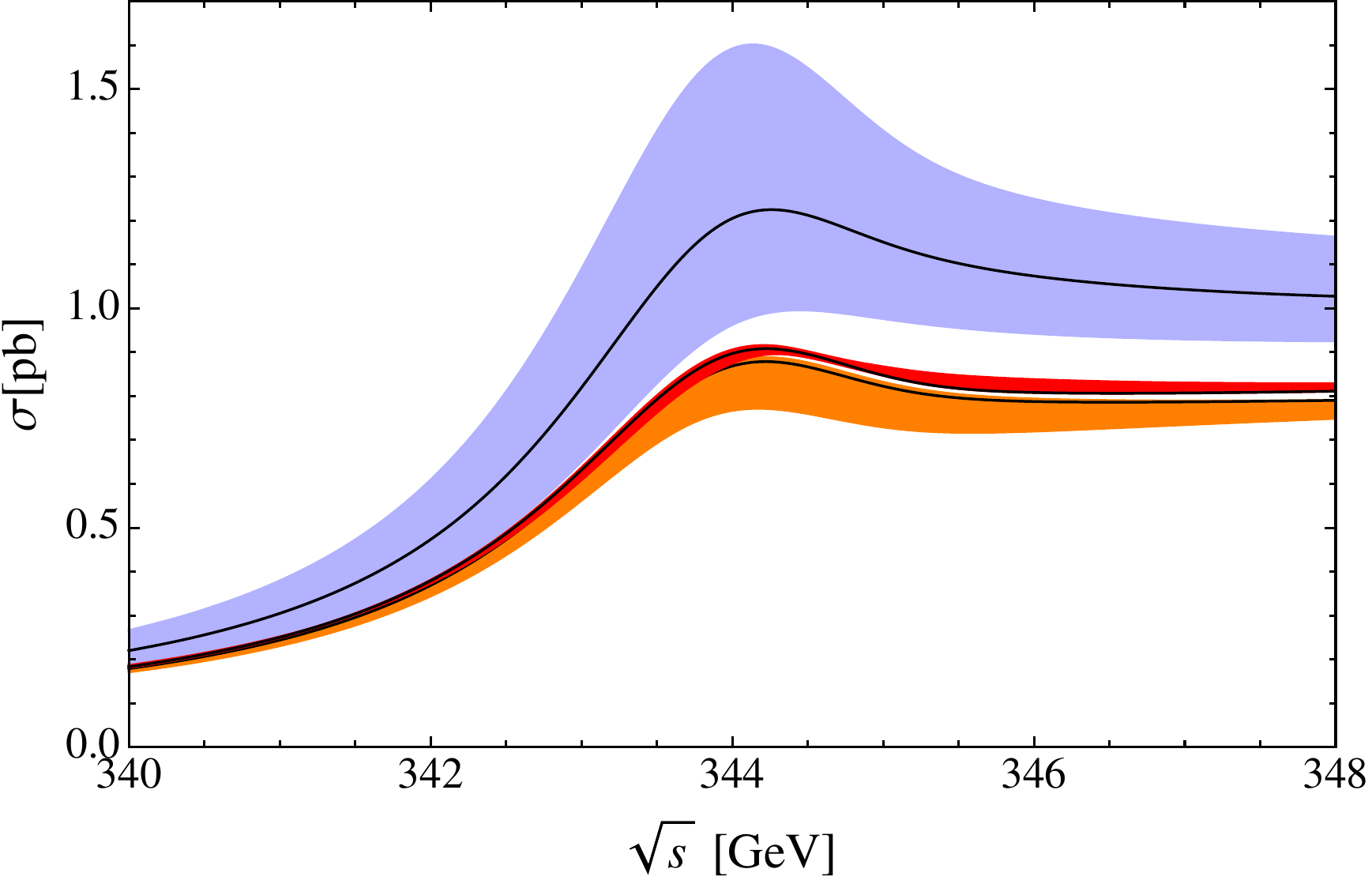}
\put(-185,120){a)}
\put(-65,111){\sf\hellblau \scriptsize LL}
\put(-65,62){\sf\orange \scriptsize NLL}
\put(-65,79){\sf\rot \scriptsize NNLL (incomp.)}
\put(-45,30){\scriptsize $\Delta M\!=\!\infty$}
\includegraphics[width=0.49\textwidth]{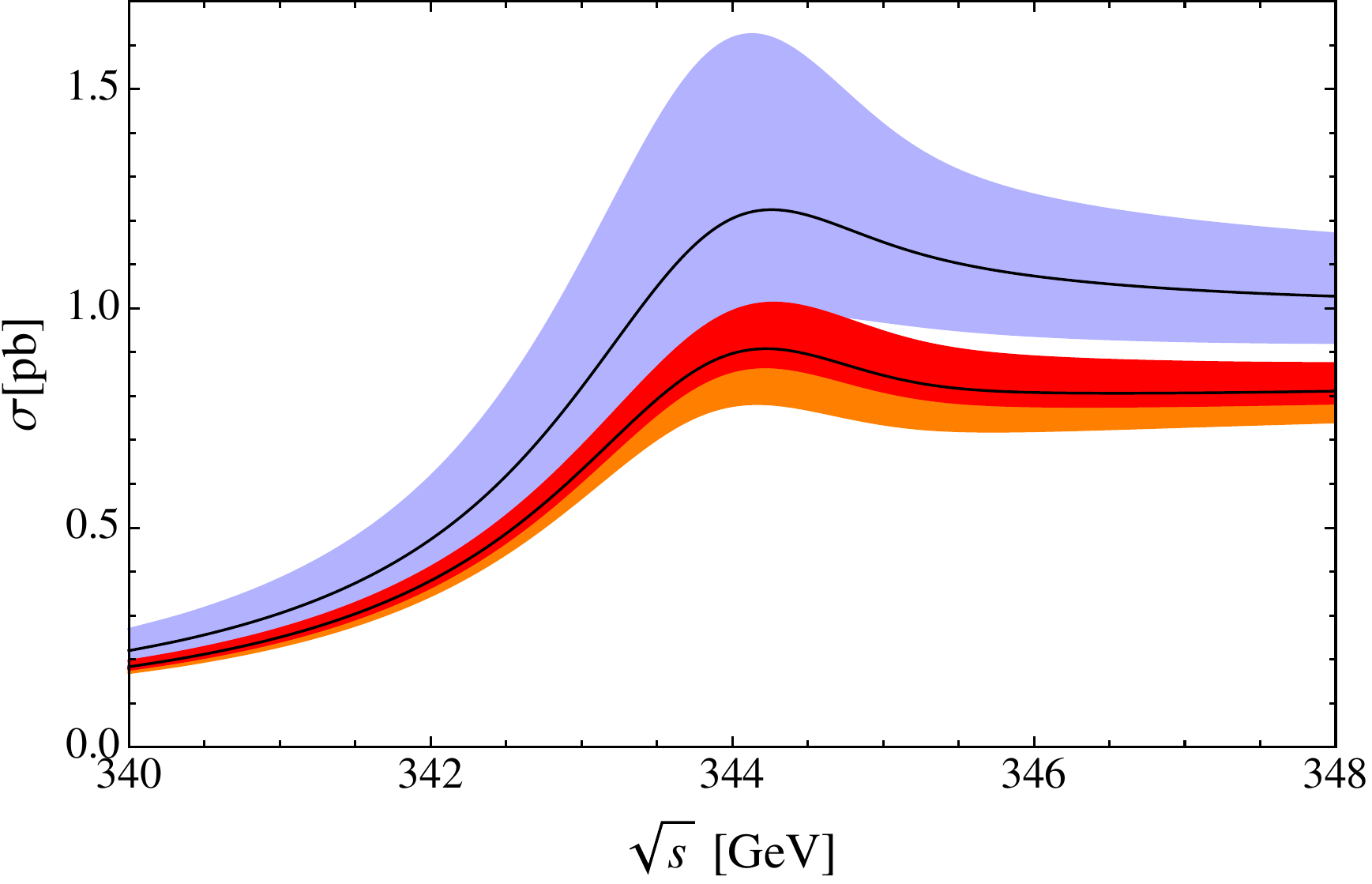}
\put(-185,120){b)}
\put(-65,112){\sf\hellblau \scriptsize LL}
\put(-65,62){\sf\orange \scriptsize NLL}
\put(-65,83){\sf\rot \scriptsize NNLL (incomp.)}
\put(-45,30){\scriptsize $\Delta M\!=\!\infty$}

\includegraphics[width=0.49\textwidth]{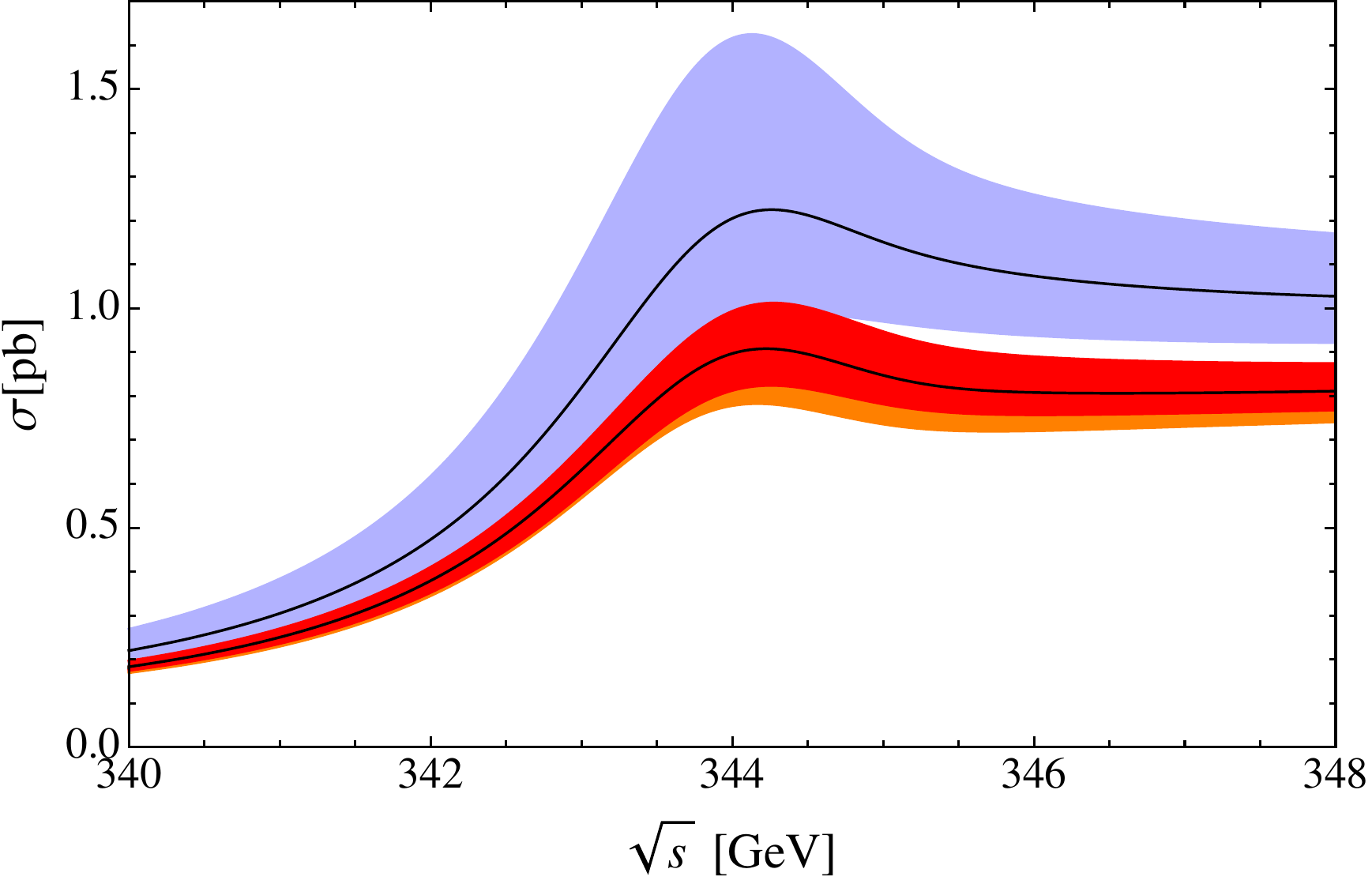}
\put(-185,120){c)}
\put(-65,112){\sf\hellblau \scriptsize LL}
\put(-65,62){\sf\orange \scriptsize NLL}
\put(-65,83){\sf\rot \scriptsize NNLL (incomp.)}
\put(-45,30){\scriptsize $\Delta M\!=\!\infty$}
\includegraphics[width=0.49\textwidth]{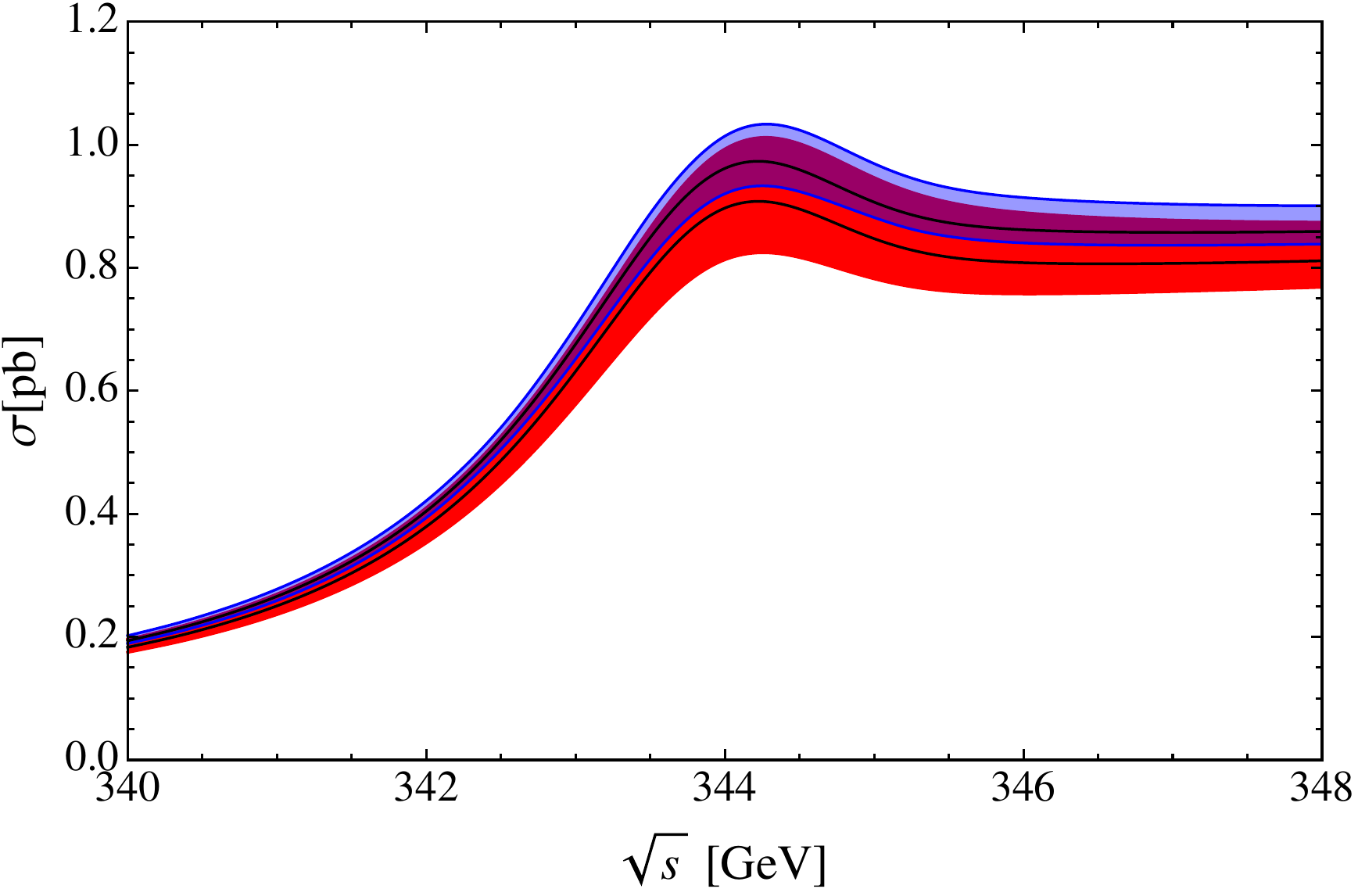}
\put(-185,120){d)}
\put(-70,86){\sf\rot \scriptsize NNLL (incomp.)}
\put(-55,110){\sf\blau \scriptsize NNLL}
\put(-45,30){\scriptsize $\Delta M\!=\!\infty$}

\caption{Plots of the inclusive ($\Delta M=\infty$) total cross section with corresponding error bands from scale variations to LL (light blue), NLL (orange) and incomplete NNLL (red) order. 
In these plots the NNLL evolution of the current coefficient $c_1$ has been switched off ($\xi^{\rm NNLL}_{\rm m}=\xi^{\rm NNLL}_{\rm nm}=0$) for comparison with previous analyses.  
The error bands in panel~a come from the variation of $f$ between 1/2 and 2 with $h=1$ fixed and in panel~b from the variation of $h$ between 1/2 and 2 with $f=1$ fixed.
Panel~c shows the scale uncertainties from combined $h$-$f$ variations scanning over the region defined in Fig.~\ref{hfregion}.
Panel~d compares the incomplete NNLL band (red) of panel~c with the new NNLL band (transparent blue) from Fig.~\ref{totcross}~c.
For the plots we used $\Gamma_t=1.5$ GeV, $M^{1\rm S}=172$ GeV and $\alpha_s^{(n_f=5)}(172~\mbox{GeV})=0.1077$.
\label{incomptotcross}}
\end{figure}
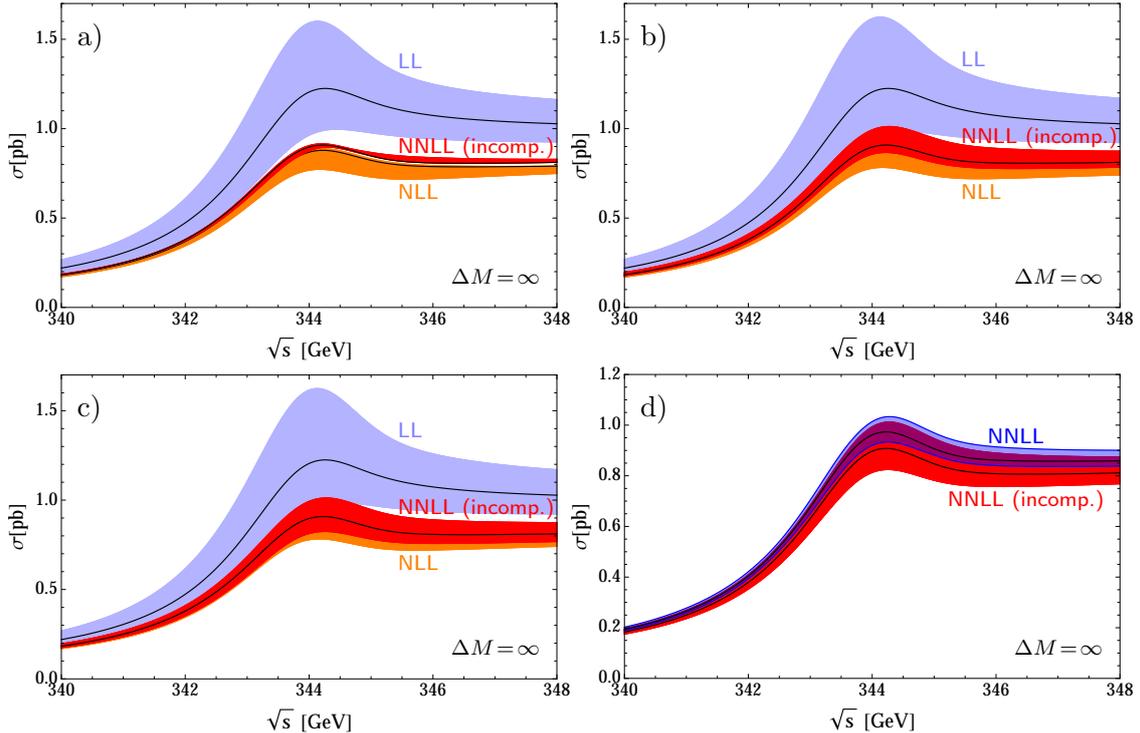

Before starting our discussion it is instructive to briefly recall the analyses and conclusions of the three previous
analyses  by  Hoang et.\ al.~\cite{Hoang:2001mm}, by Pineda and Signer~\cite{Pineda:2006ri} and Hoang~\cite{Hoang:2003xg}, 
which were based on RGI NRQCD calculations.
At the time of the analysis of Ref.~\cite{Hoang:2001mm} the NNLL anomalous dimension of the current Wilson coefficient $c_1$ were still unknown and 
the numerical examinations were carried out setting $\xi^{\rm NNLL}_m + \xi^{\rm NNLL}_{nm}=0$. 
This means that their results were missing the sizable positive correction coming from the NNLL order evolution in the square of the Wilson coefficient $c_1$ visible in the left panel of Fig.~\ref{c1XiPlot}. Hoang et al.~\cite{Hoang:2001mm} only accounted for the correlation of soft and ultrasoft renormalization scales
according to Eq.~\eqref{parametrization} with energy-independent $\nu$-values and did not vary the matching scale ($h=1$).  
They obtained a relative uncertainty for the total cross section of $\delta \sigma/\sigma \lesssim \pm 3\%$ based on variations $0.1 < \nu < 0.4$. 
We can produce this result well with our scale variation for $1/2 \le f \le 2$ setting $h=1$ and $\xi^{\rm NNLL}_m + \xi^{\rm NNLL}_{nm}=0$ as shown 
in Fig.~\ref{incomptotcross}a, exhibiting a very small relative scale variation of around $\pm 1 \%$ and a good apparent overlap between the NLL order and (incomplete) NNLL order results. 
In their subsequent analysis Pineda and Signer~\cite{Pineda:2006ri} in addition considered variations of the matching scale quite similar to our constraint 2. 
This can be reproduced by our $h$-variations, see the lines in Fig.~\ref{hfregion} indicating variations carried out in~\cite{Pineda:2006ri}.
Pineda and Signer used effectively the same incomplete NNLL QCD theory input as Hoang et.\ al.~\cite{Hoang:2001mm} and found very large matching scale dependence at the level of $\pm 10 \%$. 
Their result showed that the cross section is considerably more sensitive to global variations of the three scales $\mu_h$, $\mu_S$, $\mu_U$ (particularly when they are small) 
than to the correlated scale variation corresponding to the variation of $f$.
The same behavior is observed in Fig.~\ref{incomptotcross}b, where we have set $\nu=\nu_*$ and varied $1/2 \le h \le 2$.
giving relative variations of the cross section of around $ \pm 8\%$ at the (incomplete) NNLL order. 
For completeness we also show in Fig.~\ref{incomptotcross}c
the result from the combined $h$-$f$ variation according to the entire red area displayed in Fig.~\ref{hfregion}, causing an overall relative variation of the (incomplete) NNLL order prediction of $\pm 10\%$.

At the time of the analysis by Hoang~\cite{Hoang:2003xg}\footnote{In fact Ref.~\cite{Hoang:2003xg} appeared before Ref.~\cite{Pineda:2006ri}.} only the non-mixing corrections to the NNLL order anomalous dimension of the Wilson coefficient $c_1$ were known and his analysis was carried out setting $\xi^{\rm NNLL}_m =0$. Hoang found extremely large scale variations with very large positive shifts of the cross section due to the enormous positive size of the ultrasoft non-mixing corrections as shown in Fig.~\ref{c1XiPlot}b. 
Since the cancellation between the ultrasoft mixing and non-mixing corrections to the NNLL order anomalous dimension of $c_1$ appears to be a crucial aspect
of the behavior of the ultrasoft NNLL order corrections in the RGI cross section we do not discuss the results of Ref.~\cite{Hoang:2003xg} further.

\begin{figure}[ht]
\includegraphics[width=0.49\textwidth]{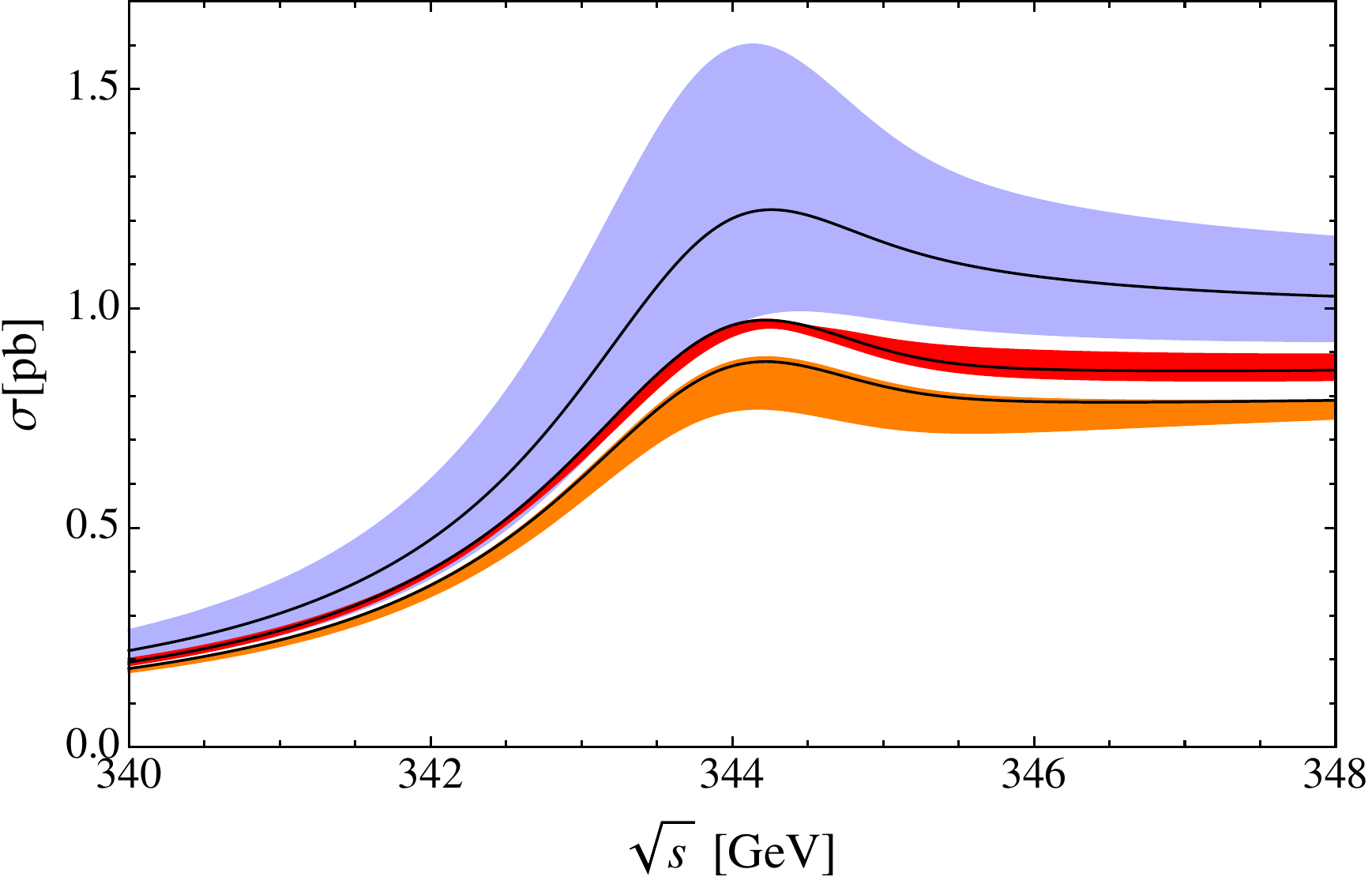}
\put(-185,120){a)}
\put(-65,111){\sf\hellblau \scriptsize LL}
\put(-65,62){\sf\orange \scriptsize NLL}
\put(-65,83){\sf\rot \scriptsize NNLL}
\put(-45,30){\scriptsize $\Delta M\!=\!\infty$}
\includegraphics[width=0.49\textwidth]{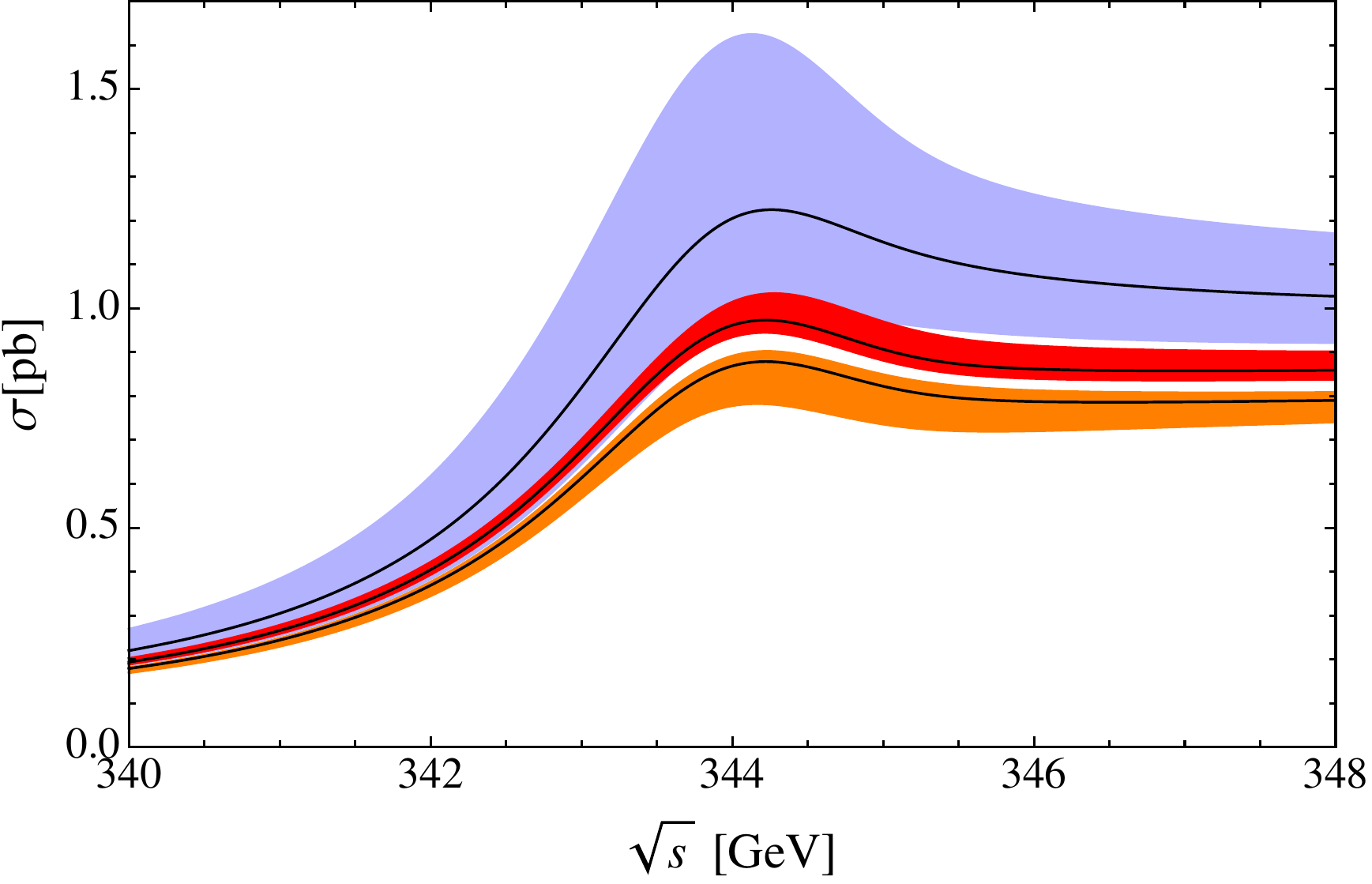}
\put(-185,120){b)}
\put(-65,112){\sf\hellblau \scriptsize LL}
\put(-65,62){\sf\orange \scriptsize NLL}
\put(-65,85){\sf\rot \scriptsize NNLL}
\put(-45,30){\scriptsize $\Delta M\!=\!\infty$}

\includegraphics[width=0.49\textwidth]{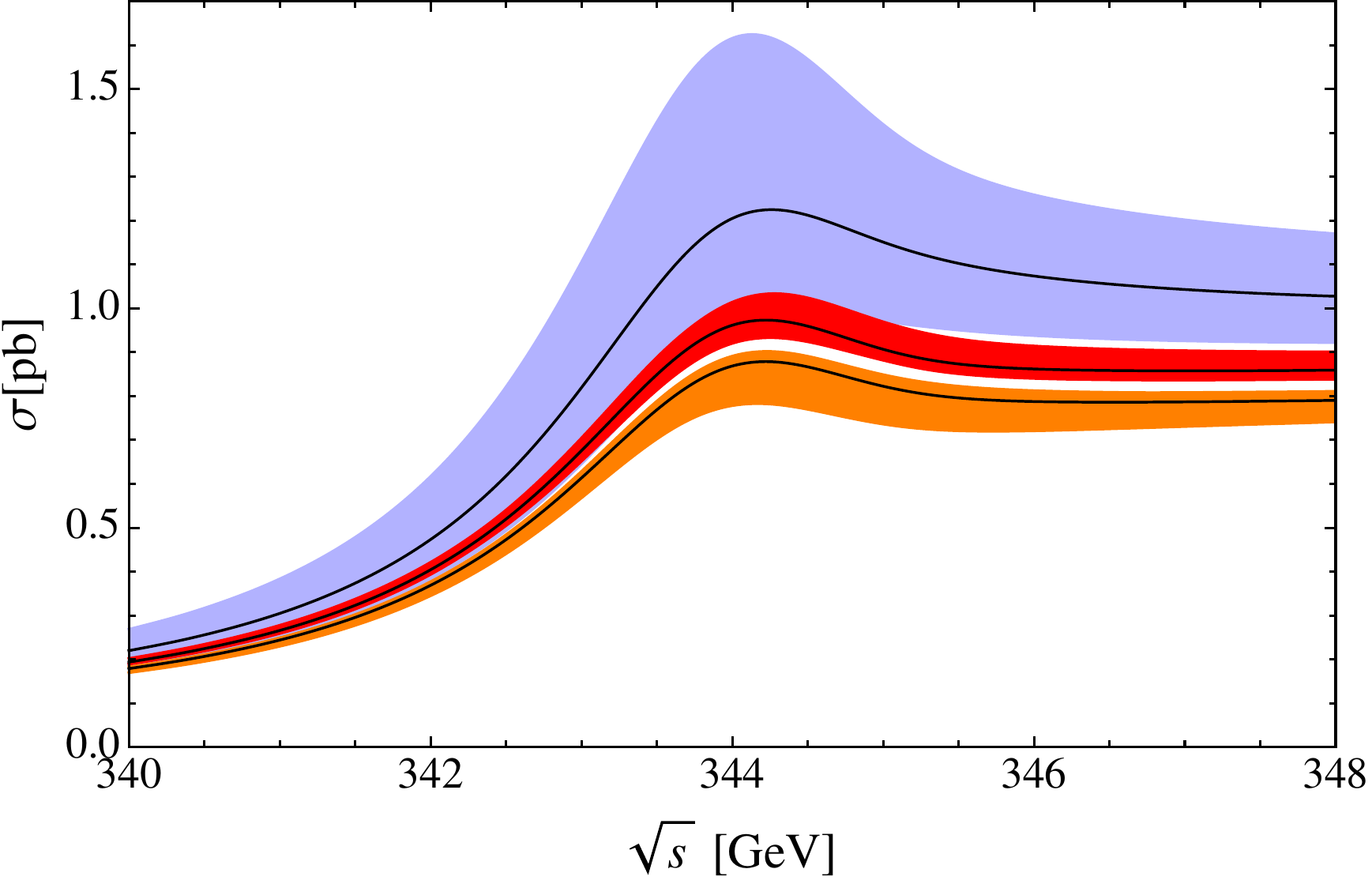}
\put(-185,120){c)}
\put(-65,112){\sf\hellblau \scriptsize LL}
\put(-65,62){\sf\orange \scriptsize NLL}
\put(-65,85){\sf\rot \scriptsize NNLL}
\put(-45,30){\scriptsize $\Delta M\!=\!\infty$}
\includegraphics[width=0.49\textwidth]{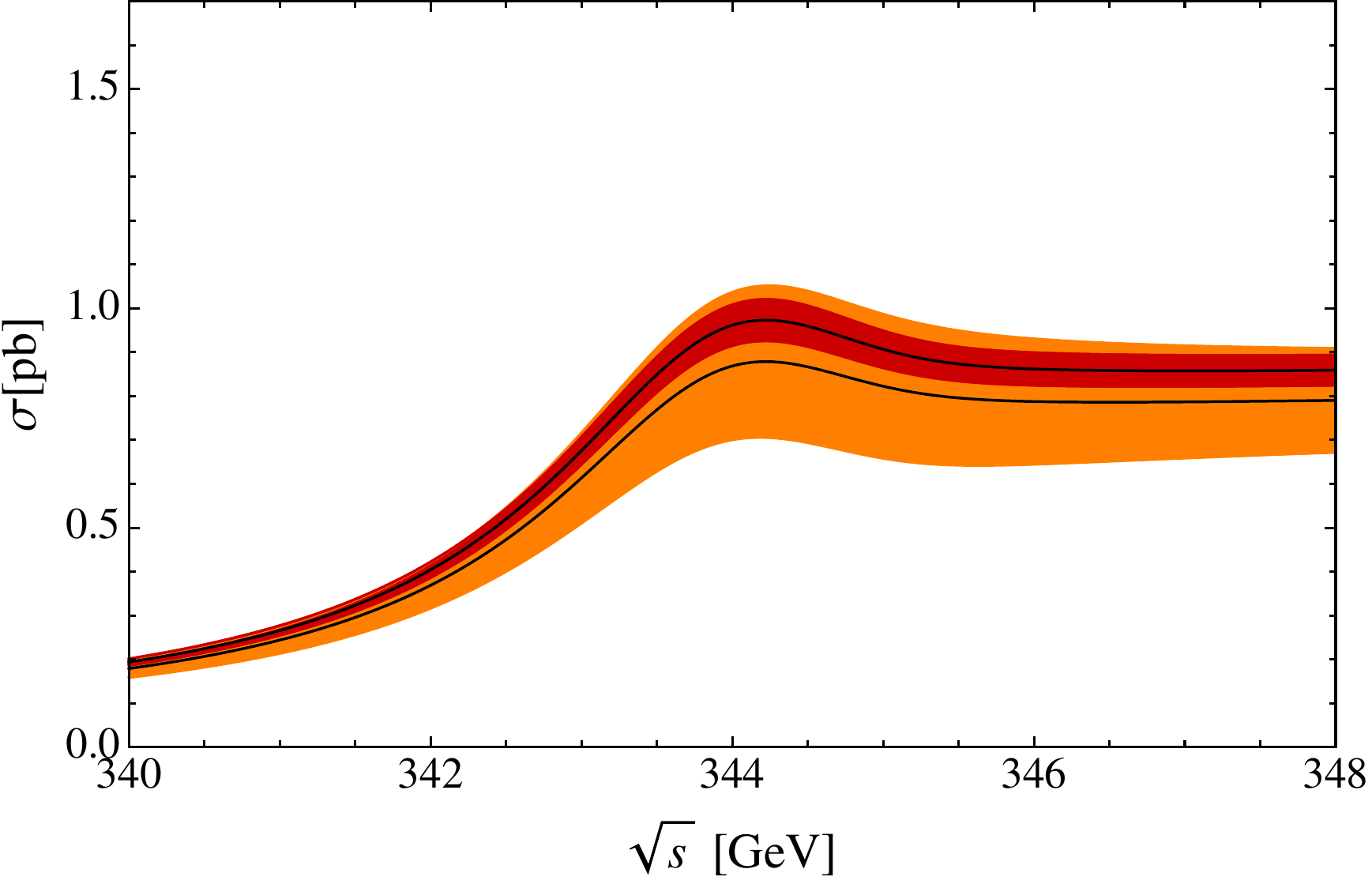}
\put(-185,120){d)}
\put(-65,57){\sf\orange \scriptsize NLL}
\put(-30,83){\sf\rot \scriptsize NNLL}
\put(-45,30){\scriptsize $\Delta M\!=\!\infty$}

\caption{
Plots of the inclusive total cross section with corresponding error bands from scale variations analogous to Fig.~\ref{incomptotcross}~a-c, but using our complete vNRQCD expression for the cross section at NNLL order.
Panel~d shows the NLL and NNLL results with the symmetric error estimated as $\pm$ half the size of the difference to the result at one lower order. While the NNLL error bands in panel~c and~d are almost identical the NLL band in panel~d is much larger than in panel~c and completely embeds the NNLL band.
The central black curves in panel~d are the same as in panel~c.
($\Gamma_t=1.5$ GeV, $M^{1\rm S}=172$ GeV, $\alpha_s^{(n_f=5)}(172~\mbox{GeV})=0.1077$)
\label{totcross}}
\end{figure}
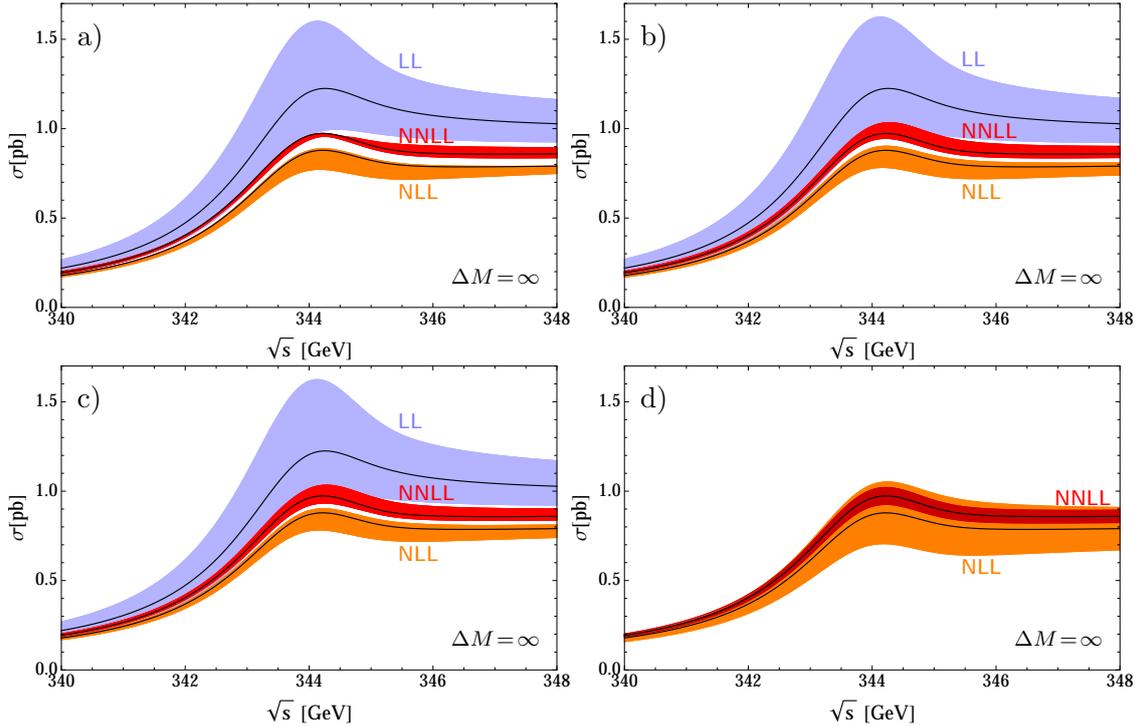

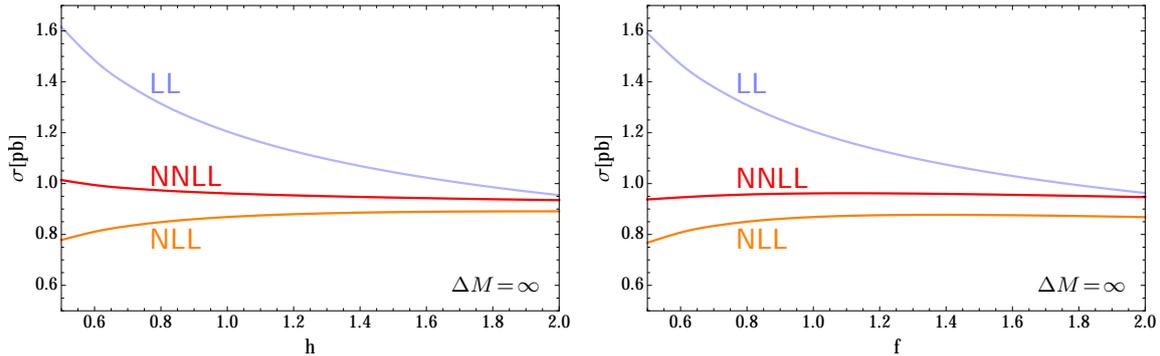
\begin{figure}[t]
\includegraphics[width=0.49\textwidth]{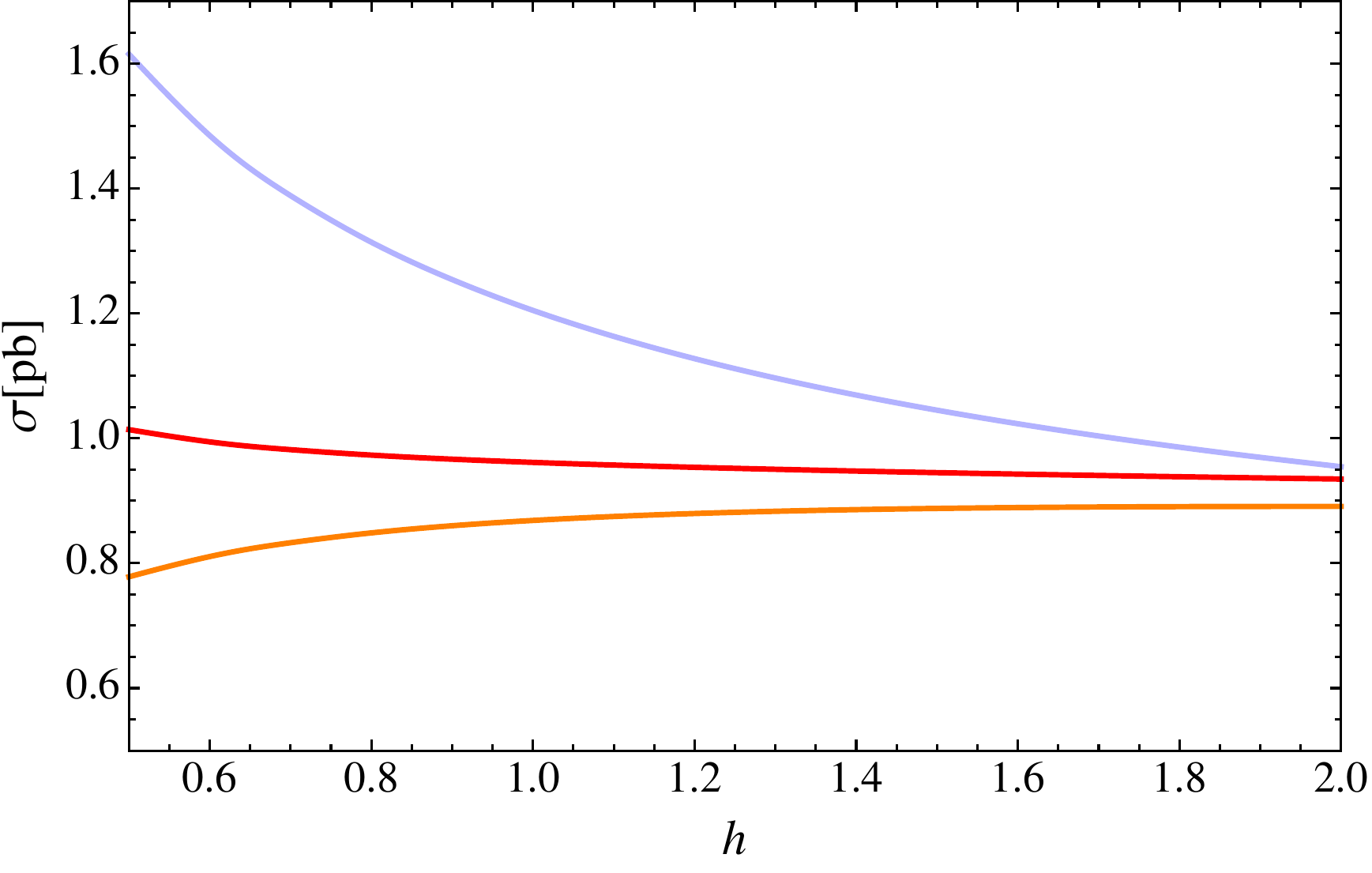}
\put(-158,101){\sf\hellblau LL}
\put(-158,42){\sf\orange NLL}
\put(-158,66){\sf\rot NNLL}
\put(-45,26){\scriptsize $\Delta M\!=\!\infty$}
\hfill
\includegraphics[width=0.49\textwidth]{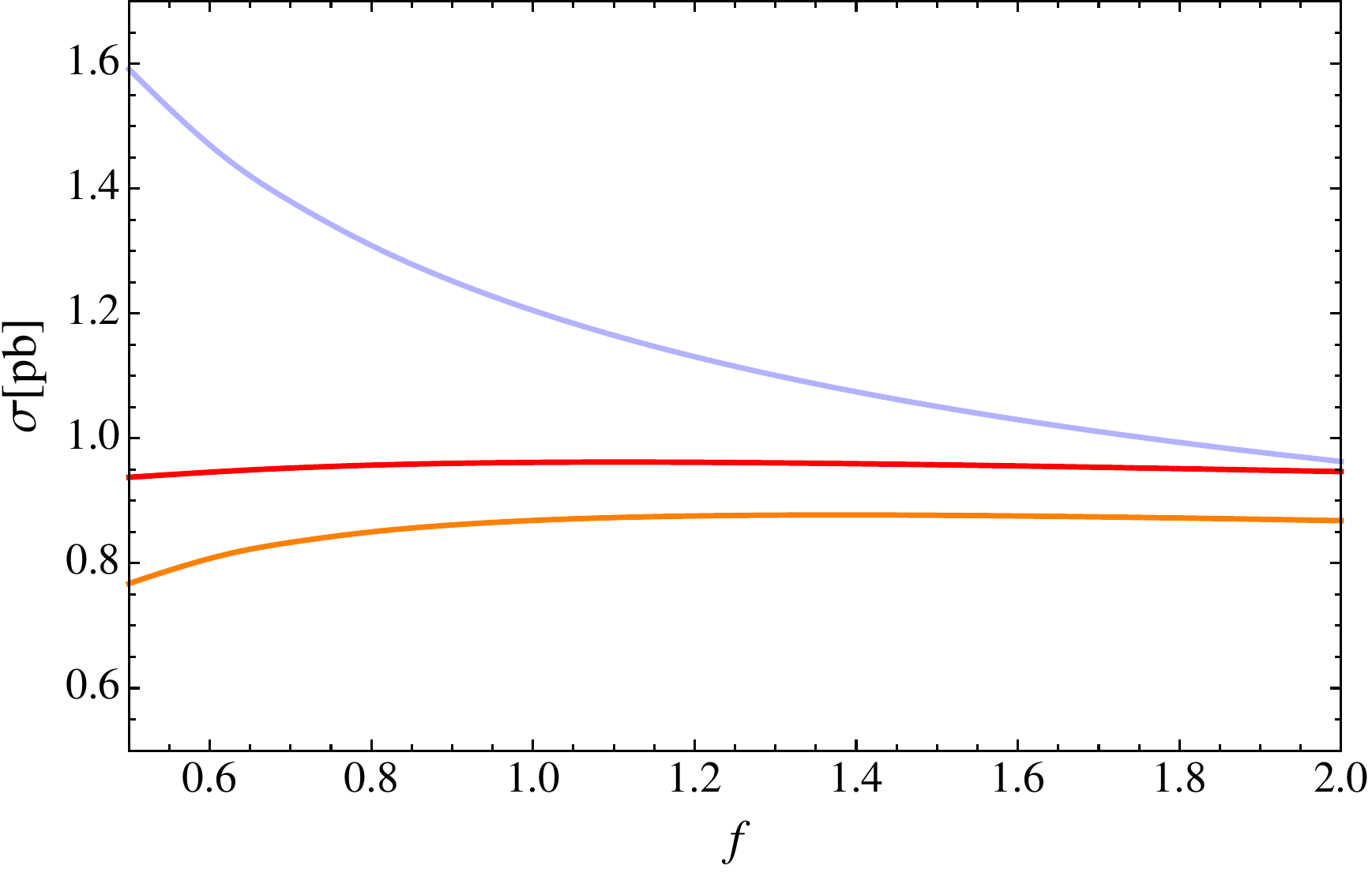}
\put(-158,101){\sf\hellblau LL}
\put(-158,42){\sf\orange NLL}
\put(-158,65){\sf\rot NNLL}
\put(-45,26){\scriptsize $\Delta M\!=\!\infty$}
\caption{$h$ and $f$ dependence of the inclusive total cross section close to the peak ($\sqrt{s}=2m$).
In the left plot we keep $f=1$ and in the right plot $h=1$ fixed.
($\Gamma_t=1.5$ GeV, $M^{1\rm S}=172$ GeV, $\alpha_s^{(n_f=5)}(172~\mbox{GeV})=0.1077$)
\label{PeakNormhfdep}}
\end{figure}

Finally, in Fig.~\ref{totcross} we show the results for the same respective scale variations as performed in Fig.~\ref{incomptotcross}, but employing our complete NNLL order vNRQCD prediction properly accounting for all known NNLL contributions to the RG evolution of $c_1$ as described in Sec.~\ref{sectioncrosssection}. 
The respective LL and NLL results are identical to those in Fig.~\ref{incomptotcross}, and the solid black lines refer to the corresponding predictions using the default values $f=h=1$. 
Since we now employ the complete NNLL order prediction, we can also discuss the convergence and consistency of the results passing from LL and NLL order to NNLL order.  
Fig.~\ref{totcross}a shows the results using only the subtraction velocity variation with $1/2 \le f \le 2$ setting $h=1$ whereas Fig.~\ref{totcross}b only displays the matching scale variation $1/2 \le h \le 2$ setting $f=1$. 
While the $f$-variation leads to an effect ($\pm 1 \%$ at the peak) on the NNLL prediction similar to Fig.~\ref{incomptotcross}a, where the incomplete NNLL order result was employed, we find that the matching scale dependence of the complete NNLL order result is reduced by about a factor of two compared to the incomplete NNLL result. 
The combined $h$-$f$-variation according to Fig.~\ref{hfregion}, which is displayed in Fig.~\ref{totcross}c, leads to a scale variation of the complete NNLL order prediction of
\begin{align}
\frac{\delta \sigma_{t\bar t}^{\rm incl.}}{\sigma_{t\bar t}^{\rm incl.}} = \pm 5\%\,.
\label{toterr}
\end{align}
In Fig.~\ref{incomptotcross}d we compare the error bands from the combined $h$-$f$-variation using the incomplete NNLL (red) and complete NNLL (transparent blue) predictions to illustrate the impact of the 
NNLL order corrections to the anomalous dimension of $c_1$.
For the interpretation of the complete NNLL scale variation, Eq.~\eqref{toterr}, it is, however, also required to analyze it in view of the LL and NLL order results. 
Considering the combined $h$-$f$ variation in Fig.~\ref{totcross}c in the peak region we find $\pm 24\%$ at LL order and $\pm 7\%$ at NLL order. 
However, there is no overlap between the LL and NLL order bands, and the NNLL order prediction is essentially covering the gap between the LL and NLL order bands. 
At this point it is instructive to also consider the LL, NLL and NNLL order curves for the default values $f=h=1$ (black lines). 
We see that at LL and NNLL order they each are well within the bands from the scale variation. 
For the NLL order prediction, however, the default prediction is very close to the upper edge of the band.
This is further visualized in Fig.~\ref{PeakNormhfdep}, where we have displayed the $h$- (left panel) and $f$-dependence (right panel) of the peak cross section at LL, NLL and NNLL order.
Given the observations it is reasonable to conclude that the NLL scale variations do not provide a good perturbative error estimate.

Alternatively one may estimate the uncertainty of the cross section at a given order by half of the separation between the default predictions at the present and the preceding order.
This method gives $20\%$ at NLL order and $5\%$ at NNLL order (at the peak position) as can be read off from the bands in Fig.~\ref{totcross}d.
The NLL uncertainty is almost three times larger than the one from scale variations, while at NNLL order both methods yield the same uncertainty.
Furthermore, replacing the NLL error band in Fig.~\ref{totcross}c with the one of Fig.~\ref{totcross}d leads to a perfectly consistent
perturbative behavior of the LL, NLL and NNLL order predictions at all energies in the threshold region. 
We therefore conclude that the NNLL perturbative error given in Eq.~\eqref{toterr} is reliable.

\begin{figure}[ht]
\includegraphics[width=0.49\textwidth]{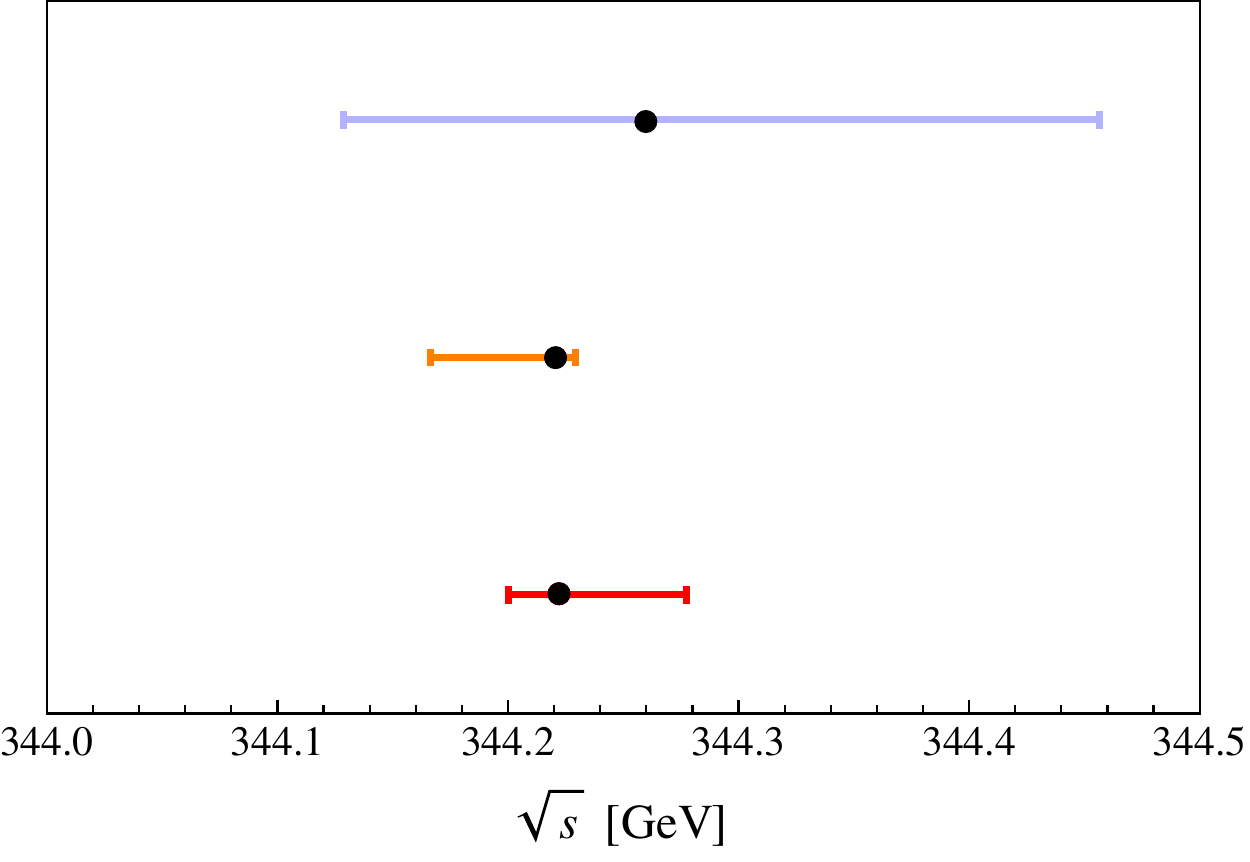}
\put(-176,122){\sf\hellblau LL}
\put(-176,82){\sf\orange NLL}
\put(-176,42){\sf\rot NNLL}
\put(-196,132){a)}
\put(-48,32){\scriptsize $\Delta M\!=\!\infty$}
\hfill
\includegraphics[width=0.49\textwidth]{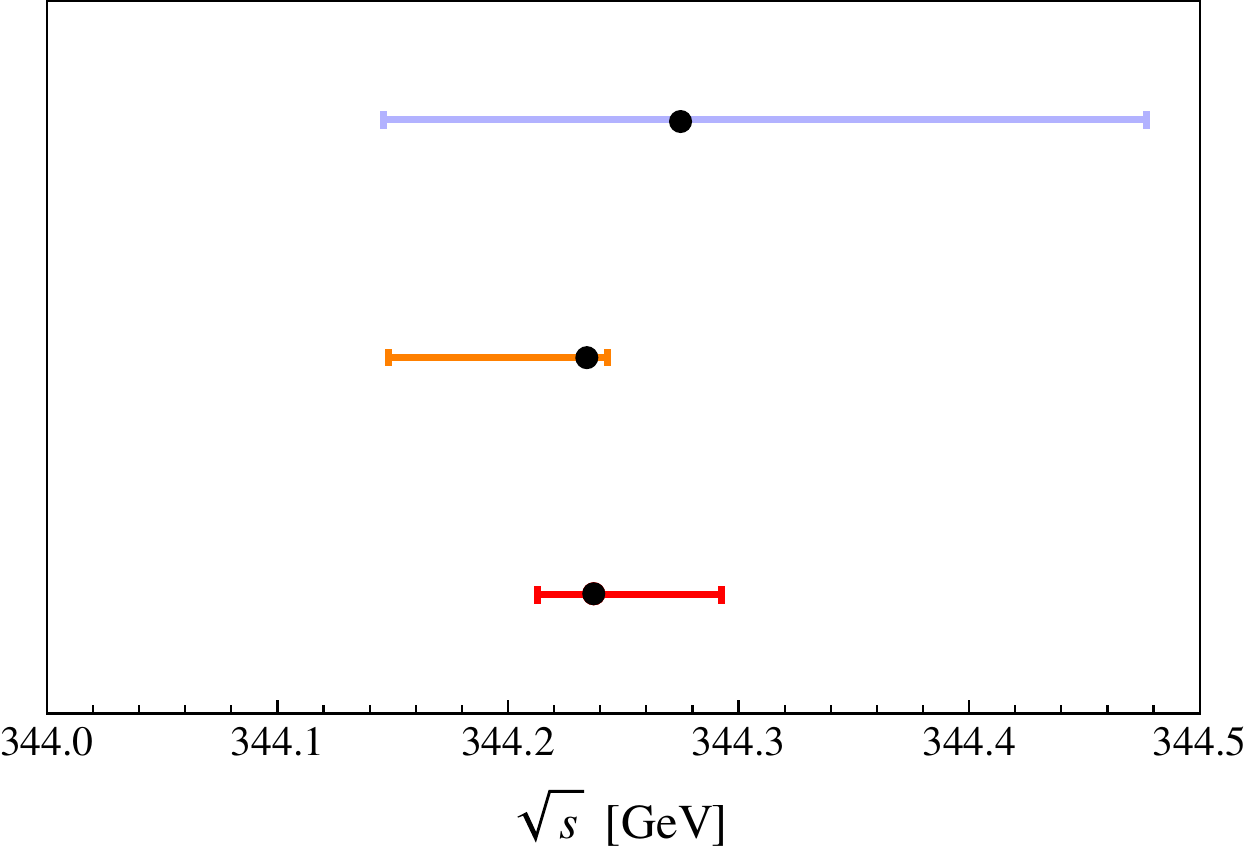}
\put(-176,122){\sf\hellblau LL}
\put(-176,82){\sf\orange NLL}
\put(-176,42){\sf\rot NNLL}
\put(-196,132){b)}
\put(-64,32){\scriptsize $\Delta M\!=\! 30$ GeV}
\caption{Resonance peak position of the inclusive total cross section (a) and the cross section with a top/antitop invariant mass cut ($\Delta M=30$ GeV) (b) at LL, NLL and (full) NNLL precision for $h=f=1$ (black dots). 
The error bars indicate the associated uncertainties from combined $h$-$f$ scale variations.
($\Gamma_t=1.5$ GeV, $M^{1\rm S}=172$ GeV, $\alpha_s^{(n_f=5)}(172~\mbox{GeV})=0.1077$)
\label{peakpos}} 
\end{figure}
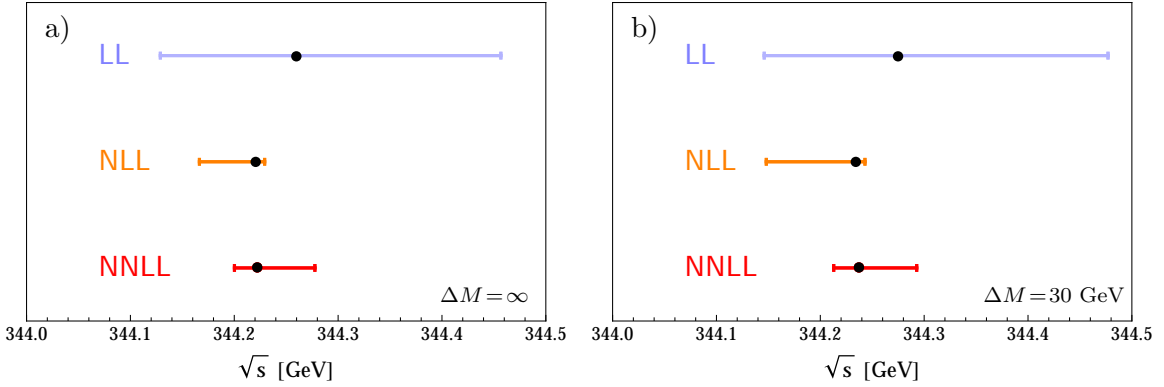

It is also interesting to study the dependence of the peak position on the scale variations. 
In Fig.~\ref{peakpos}a we have displayed the range of variation (from scaning the $h$-$f$ region shown in Fig.~\ref{hfregion}) of the peak position at LL, NLL and NNLL order. 
The respective black dots indicate the peak positions for the default scales $h=f=1$. 
Similar as for the line-shape shown in Fig.~\ref{totcross}c, we find that the NLL and NNLL scale variations are
much smaller than the LL order one, but there does not seem to be any considerable improvement at NNLL order compared to the NLL result. 
The location of the default peak position, on the other hand, 
reveals that the NLL order scale dependence is again single-sided, and we therefore conclude that the NLL scale variation underestimates the perturbative error, just as it does for the normalization. 
Likewise we believe that the NNLL order variation range, which is about 80 MeV provides a reliable estimate of the NNLL perturbative uncertainty. 
Given that the peak position is related to twice the 
top quark mass this indicates that the theoretical uncertainty in a top quark mass determination from a threshold scan is
\begin{align}
\delta m \sim \pm 20 {\rm MeV}\,. 
\label{deltamass}
\end{align}
We stress, however, that eventually the top quark mass will be measured from fits to the experimental line-shape measurements, which involves on the theoretical side a convolution with the $e^+e^-$ luminosity spectrum accounting for effects such as the beam strahlung and initial state radiation. 
So the result in Eq.~\eqref{deltamass} can only serve as a first, naive, error estimate. 
A reliable method should be based on scale variations of the full theory code used for the fits and should be carried out during the analysis of the experimental data.

\section{Phase Space Cuts}
\label{sectioncuts} 

We now analyze our results including a cut in the form of Eq.~\eqref{pscut} on the reconstructed top and antitop invariant masses using the approach explained in Sec.~\ref{sectioncrosssection}.
From the physical point of view the invariant mass cut removes unphysical contributions in the NRQCD prediction directly related to the top width implementation in Eq.~\eqref{veff} and the use of the optical theorem~\eqref{effRratios} - the method all previous analyses of the top pair threshold cross section relied on.
The contributions from invariant masses above the cut are mainly unphysical because they are calculated with the nonrelativistic approximation for which the phase space allows for arbitrary large invariant masses.
As was emphasized in Ref.~\cite{Hoang:2010gu}, these unphysical terms are positive and, in fact, are the reason why the total top threshold cross section based on Eqs.~\eqref{effRratios} and~\eqref{veff}
never vanishes even for energies far below the top pair threshold. 

While it is the purpose of the phase space matching procedure advocated in Ref.~\cite{Hoang:2010gu} to get rid of these unphysical contributions and implement the correct 
behavior of off-shell configurations from the underlying electroweak theory, 
the phase space cuts are special because the unphysical contributions are the by far dominating contribution in the phase space matching~\cite{Hoang:2010gu}. 
It is therefore interesting to have a closer look into the perturbative behavior of the total cross section after an invariant mass cut has been imposed. 
Since the perturbative behavior of highly virtual unphysical NRQCD phase space configurations has never been analyzed before, it is instructive to examine the convergence and the scale-dependence of the cross section with the invariant mass cut. 
Moreover, as was demonstrated in Ref.~\cite{Hoang:2010gu}, the cross section with the invariant mass cut is numerically much closer to the true inclusive total cross section (in the full electroweak theory).

\begin{figure}[t]
\includegraphics[width=0.494\textwidth]{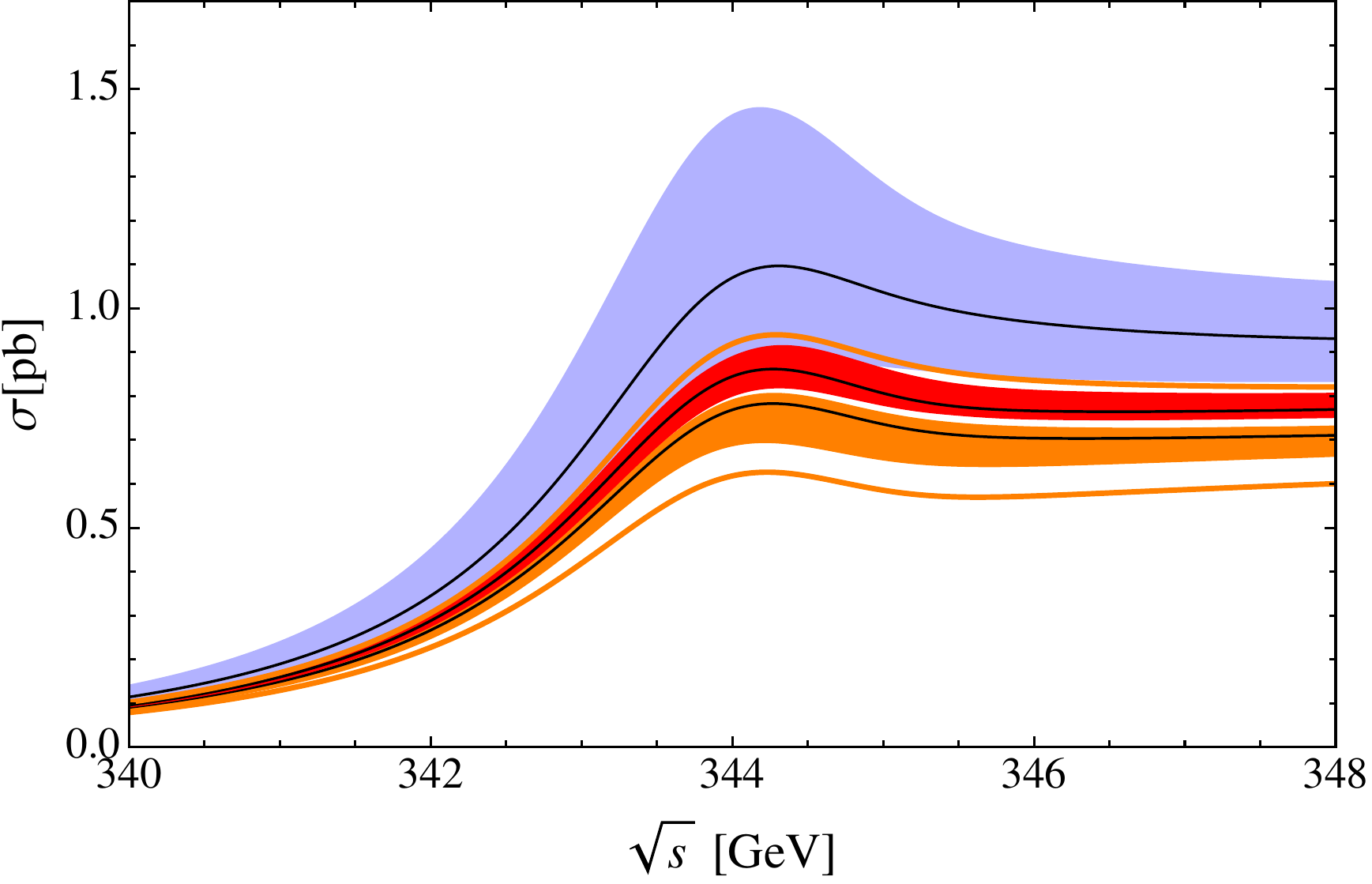}
\put(-186,123){a)}
\put(-63,30){\scriptsize $\Delta M\!=\! 15$ GeV}
\put(-38,98){\sf \scriptsize\hellblau LL}
\put(-70,52){\sf \scriptsize\orange NLL}
\put(-44,78){\sf \scriptsize\rot NNLL}
\hfill
\includegraphics[width=0.494\textwidth]{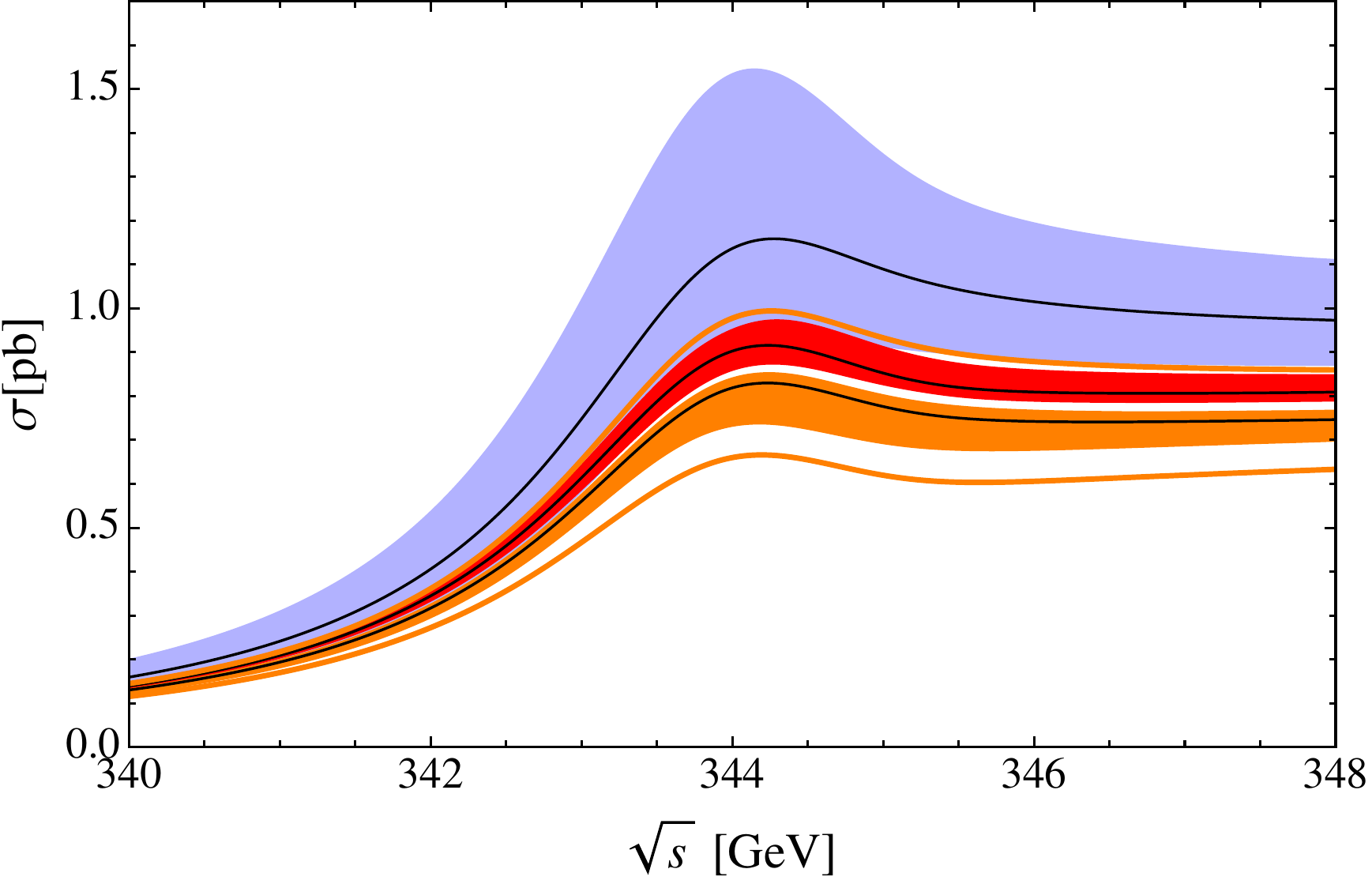}
\put(-186,123){b)}
\put(-63,30){\scriptsize $\Delta M\!=\! 30$ GeV}
\put(-38,102){\sf \scriptsize \hellblau LL}
\put(-70,54){\sf \scriptsize \orange NLL}
\put(-44,81){\sf \scriptsize \rot NNLL}
\caption{Cross section with a cut of $\Delta M=15$~GeV (a) and $\Delta M=30$~GeV (b) on the invariant masses of the reconstructed top and antitop. The colored bands represent the uncertainties from combined $h$-$f$ scale variations analogous to Fig.~\ref{totcross}c. Default results are indicated by the black solid lines (top: LL, middle: NNLL, bottom: NLL).
 In addition we display the alternative error estimate at NLL based on the difference between the LL and NLL default curves (band between the orange lines).
($\Gamma_t=1.5$ GeV, $M^{1\rm S}=172$ GeV, $\alpha_s^{(n_f=5)}(172~\mbox{GeV})=0.1077$)
\label{cutcross}} 
\end{figure}
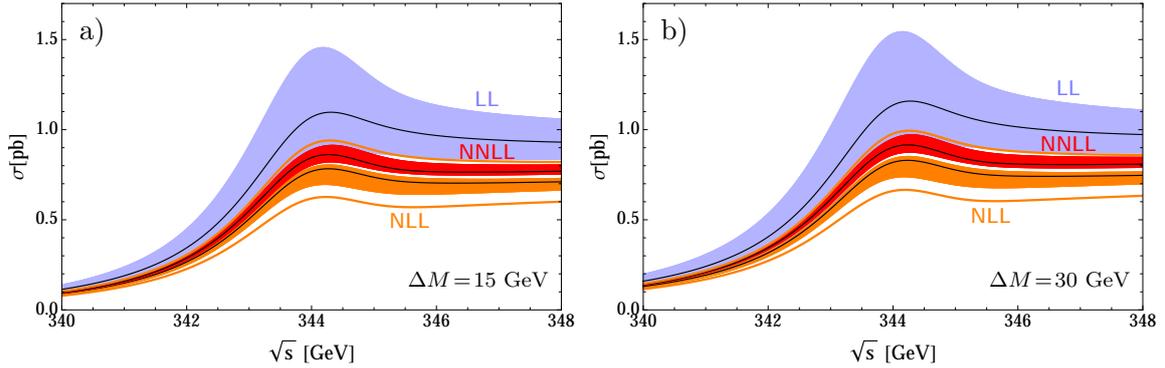

In Fig.~\ref{cutcross} we have plotted the LL, NLL and NNLL order cross sections with an invariant mass cut $\Delta M=15$ GeV (left panel) and $\Delta M=30$ GeV (right panel).  
The uncertainty bands are obtained by the combined $h$-$f$-variations indicated in Fig.~\ref{hfregion}. 
The respective black lines are the predictions with the default scale choice $f=h=1$. 
Moreover we have indicated the (symmetric) NLL error estimate from the separation of the LL and NLL default predictions as additional orange lines.
Overall we see that the behavior of the LL, NLL and NNLL order predictions concerning the scale variation and the convergence properties is very similar to the predictions for $\Delta M=\infty$ discussed in Sec.~\ref{sectionanalysis}. 
As we have already concluded for the $\Delta M=\infty$ case at NLL order the scale variation certainly underestimates the theoretical uncertainty, and one should rely on the method based on the default predictions.  
At NNLL order the uncertainty estimates coming from scale variation and the method based on the default predictions are equivalent and give 
\begin{align}
\frac{\delta \sigma_{t\bar t}^{\rm cut}}{\sigma_{t\bar t}^{\rm cut}} = \pm 5\%\,.
\label{cuterr}
\end{align}
We consider this estimate reliable as discussed above in Sec.~\ref{sectionanalysis}. 
Since the invariant mass cuts imposed here represent the largest contribution in a complete treatment of
electroweak effects~\cite{Hoang:2010gu} we can further conclude that this theoretical uncertainty also applies to the NNLL order prediction with the full set of electroweak corrections. 
We emphasize that our prediction here only includes the effects of the width and of the invariant mass cut, and that some of the electroweak corrections we neglected in this analysis lead to changes of the line-shape that must be taken into account for data analysis.

We conclude the discussion of the cross section by mentioning that the invariant mass cut has a considerable numerical effect on the cross section for c.m.\ energies below the threshold peak. 
Because the unphysical phase space contribution technically behaves like a background~\cite{Hoang:2010gu}, its energy dependence is very small.
For $\Delta M=15 (30)$ MeV the phase space cuts reduce the cross section by about $0.1 (0.05)$ pb, which represents an order unity effect a few GeV below the peak position.
Physically the invariant mass cut ensures that the cross section properly vanishes at the c.m.\ energy $2(m-\Delta M)$.
We stress that the NRQCD cross section based on the optical theorem through Eqs.~\eqref{effRratios} and~\eqref{veff} and computed without phase space cut is non-vanishing for all energies, so that the predictions for the line-shape below
the peak position are unreliable. This should be taken into account in present simulation studies, in particular in connection with conclusions concerning background
and if the outcome of the fits has a significant dependence on the energy region below the peak.
 
Finally, we also comment on the peak position for the cross section with the invariant mass cut. 
In Fig.~\ref{peakpos}b we show the scale variations (based on the combined $h$-$f$ variations 
indicated in Fig.~\ref{hfregion}) of the peak position at LL, NLL and NNLL order. 
Comparing to the results for $\Delta M=\infty$ we see that the outcome concerning convergence and scale variations is essentially unchanged and that our discussion on the peak position in Sec.~\ref{sectionanalysis} is not affected by imposing a moderate phase space cut $\Delta M \gsim 20$ GeV.  
The phase space cuts also do not have any significant effect on the peak position due to their small energy-dependence. We note, however, that some of the electroweak corrections we have neglected in this analysis, most notable the interference terms, can shift the peak position at the level of 30 to 50 MeV~\cite{Hoang:2004tg}. 
So electroweak effects are highly important and cannot be neglected for experimental analyses.

\section{Conclusion}
\label{sectionconclusion}

In this paper we have analyzed for the first time the NNLL RGI top-antitop threshold cross section in $e^+e^-$ collisions with the full set of available NNLL QCD corrections in RGI perturbation theory. 
Our theory code implements the latest results on the NNLL order corrections to the anomalous dimension of the dominant top pair production current.
At this time, at NNLL order concerning QCD corrections, there is only one set of NNLL soft ("mixing") corrections missing  which is related to the ignorance of the soft NLL RG evolution of the subleading (non-Coulomb) QCD potentials. 
We presented arguments that show that these corrections are very likely tiny and negligible, so that our NNLL QCD prediction can be considered complete for practical purposes. 
We find that the NNLL QCD top pair cross section has a remaining relative theory error of $\delta \sigma/\sigma =\pm 5\%$ in the peak region and above. 
The uncertainties concerning the peak position indicate that the theoretical uncertainty in the determination of the top quark mass (in a proper threshold short-distance mass scheme) is at the level of 20 MeV. 
A reliable theory uncertainty estimate in the top mass measurement can, however, only be obtained from a simulation study including realistic experimental input, in particular related to the $e^+e^-$ luminosity spectrum, which depends on the linear collider design. 

In our analysis we did not account for the full set of presently available electroweak corrections and only included the effects from the top quark decay width and from cuts on the reconstructed top and antitop invariant masses. 
These two effects are the numerically largest sources of electroweak corrections that affect the normalization of the cross section, so that our conclusions concerning the (QCD) uncertainties remain unchanged when all electroweak corrections are taken into account.
We emphasize that the implementation of invariant mass cuts is essential to achieve realistic predictions for the cross section in the energy range below the peak, where the cross section is small.

 \acknowledgments{
 We would like to thank the Erwin Schr\"odinger Institute program "Jets and Quantum Fields for LHC and Future Colliders" for hospitality while portions of this work were completed.
 We are grateful to Thomas Teubner for his help concerning the TOPPIK program.
 MS would like to thank the particle physics theory group of the University of Vienna for hospitality and Pedro Ruiz-Femen\'ia for useful discussions on the electroweak effects.
 This work was supported in part by the DFG under Emmy-Noether Grant No. TA 867/1-1.
 }

\appendix

\section{N$^3$LO current logarithms} \label{secN3LOc1}

Expanding the NNLL RG evolution factor of the current coefficient $c_1$ in $\ah\equiv \alpha_s(h m)$ up to $\ord(\ah^3)$, which corresponds to N$^3$LO in the fixed-order counting, we obtain the expression
\begin{align}
\label{c1N3LO}
&\frac{c_1(\nu,h)}{c_1(1,h)}=
1 \;+\; \frac{1}{6} \,\ah^2\, C_F \big[2 C_F (\bmS^2\!-\!3)-3 C_A \big]\ln \nu \;+ \\
&\quad\ah^3 \Bigg\{\ln \nu \bigg[\frac{\beta_1 C_F \big[2 C_F (\bmS^2-3)-3 C_A \big]}{24 \pi 
   \beta_0}+\frac{\beta_0 C_F \big[8 C_F (\bmS^2-12)-111 C_A \big]}{288 \pi } \nn\\
&\qquad-\frac{C_F \big[5 C_A^2 (576 \ln 2 - 271)-40 C_A C_F (2 \bmS^2-75-72 \ln 2)\big]}{1440 \pi} \nn\\
&\qquad-\frac{C_F^2 \big[5 C_F (3\bmS^2+46-72 \ln 2)+6 T_F \big]}{90 \pi}
-\frac{C_F \ln h \big[8 C_A^2+C_A C_F (41-7 \bmS^2)+18C_F^2 \big]}{12 \pi} \bigg] \nn\\
&\qquad+\ln ^2\nu \bigg[\frac{\beta_0 C_F \big[3 C_A-2 C_F (\bmS^2-3) \big]}{12 \pi }
-\frac{C_F \big[8 C_A^2+C_A C_F (41-7 \bmS^2)+18 C_F^2 \big]}{24 \pi }\bigg]
\Bigg\} \nn\\
&\quad+\ord(\ah^4)\,,\nn   
\end{align}
where $\bmS^2=2$ for the spin triplet state and $\beta_0 = \frac{11}{3}C_A - \frac43 n_f T_F$, $\beta_1 = \frac{34}{3}C_A^2 - 4 C_F\, n_f T_F - \frac{20}{3}C_A n_f T_F$.

\bibliographystyle{JHEP}
\bibliography{MeittbarBib}

\end{document}